\documentclass[a4paper, 11pt]{article}
\usepackage{natbib}
\usepackage[english]{babel}

\usepackage[hmargin = 2.54cm, vmargin = 2.54cm]{geometry}
\newcommand{\utwi}[1]{\mbox{\boldmath $ #1$}}

\usepackage{booktabs,dcolumn}
\usepackage{psfrag,graphicx}
\usepackage{eso-pic,graphicx}
\usepackage{amsmath}
\usepackage{graphicx}
\usepackage{pdflscape}
\usepackage{longtable}
\usepackage{epstopdf}
\usepackage{appendix}
\usepackage{dashbox}
\usepackage{algorithm2e}
\usepackage{titlesec}
\usepackage{etoolbox}
\usepackage{setspace}
\usepackage[colorlinks=true,allcolors=blue]{hyperref}

\usepackage[export]{adjustbox}
\usepackage{caption}
\usepackage{xcolor,colortbl}

\definecolor{verylightgray}{gray}{0.9}
\newcommand{\cg}{\cellcolor{verylightgray}}
\newcolumntype{z}[1]{D{.}{.}{#1}}

\newcommand{\ubeta}{\utwi{\beta}}

\date{}

\begin{document}

\title{
\begin{center}
{\Large \bf Combining a Large Pool of Forecasts of Value-at-Risk and Expected Shortfall}
\end{center}
}
\author{James W. Taylor$^{1}$\footnote{Email: james.taylor@sbs.ox.ac.uk.}, Chao Wang$^{2}$\footnote{Corresponding author. Email: chao.wang@sydney.edu.au.}
\\
$^{1}$ Sa\"id Business School, University of Oxford, \\
$^{2}$ Business School, The University of Sydney}
\date{} \maketitle
\date{} \maketitle

\begin{abstract}
\noindent
\vspace{0.5cm}

We consider the combination of value-at-risk (VaR) and expected shortfall (ES) forecasts when a large pool of candidate forecasts is available. Given the limited literature in this area, we implement a variety of new combining methods. In terms of simplistic methods, in addition to the mean, we consider the median and mode. As a complement to the previously proposed performance-based weighted combinations, we use regularisation to reduce overfitting in the presence of many weights. Treating VaR and ES forecasts jointly as interval forecasts allows the application of adapted interval forecast combination methods, including trimmed means and a mixtures approach based on inferred probability distributions. In an empirical study involving 90 forecasting methods, trimmed mean combinations, the mixtures method, and performance-based weighting delivered particularly strong results. However, greater forecasting accuracy resulted for a pool of just six methods, chosen to ensure diversity, with performance-based weighting producing the best overall performance.\\

\noindent {\it Keywords}: Value-at-Risk; Expected Shortfall; Forecast Combining.

\end{abstract}

\titlespacing*{\section}{0pt}{*0}{*0}
\titlespacing*{\subsection}{0pt}{*0}{*0}
\titlespacing*{\subsubsection}{0pt}{*0}{*0}

\newpage
\vspace{-3em}
\section{Introduction}\label{introduction_sec}

Financial risk management is an essential task for financial institutions. A widely-used risk measure is value-at-risk (VaR), which is typically reported for daily returns on an institution’s portfolio. The VaR is a conditional quantile in the lower tail of the probability distribution of the returns. Although the VaR has the appeal of being straightforward to interpret, a limitation is that it conveys no information regarding exceedances beyond the VaR. In this regard, a better measure of risk is the expected shortfall (ES), which is the conditional expectation of the exceedances. Further advantages of ES are that, unlike VaR, ES is a coherent risk measure \citep{acerbi2002spectral} and it is subadditive, meaning that the measure for a portfolio is no greater than the sum of the measures for the constituent parts of the portfolio \citep{artzner1999}. The estimation of the ES for a 2.5\% probability level became widespread following the Basel III Accord \citep{lazar2019model}, which emphasised the use of ES in its banking regulatory framework.

Since the introduction of ES to the Basel III recommendations, the sizeable literature on VaR forecasting has been complimented by a growing variety of methods for predicting ES. These have included nonparametric methods employing simplistic approaches, such as historical simulation, and fully parametric approaches, based, for example, on a variety of GARCH models with alternative distributional assumptions. An important development in this area is the work of \cite{Fissler2016}, who show that, although ES is not elicitable by itself, it is jointly elicitable with the VaR (see also \citealp{acerbi2002coherence}). They introduce a class of joint scoring functions for VaR and ES, which has enabled the direct modelling of the VaR and ES. These include the models of \cite{taylor2019al} that are based on the CAViaR models of \cite{caviar}. The variety of methods available has been expanded by the inclusion of realised volatility and intraday range  (see, e.g., \citealp{gerlach2020semi}).

When multiple candidate forecasting methods are available, an alternative to method selection is forecast combination. The appeal of this is that it enables the synthesis of potentially different information provided by the different forecasts \citep{kang2022forecast}. Furthermore, a combination of forecasts can be viewed as a portfolio that diversifies the risk of selecting a poor forecast. Although there are large literatures on combining for point forecasts and density forecasts, little has been written on combining for VaR and ES forecasts. Indeed, the impressive review of forecast combining by \cite{wang2023forecast} makes no mention of VaR and ES.

Nevertheless, there have been several combining papers in this area. \cite{taylor2020forecast} proposes two different methods involving performance-based weighted combinations estimated using one of the joint scores of \cite{Fissler2016}. Empirical illustration is provided for VaR and ES forecasts from five methods. \cite{trucios2023comparison} use the same methods to combine forecasts from 10 methods. For the combination of 15 GAS models, \cite{wang2022forecasting} consider the mean, median, and trimmed mean, with trimming percentage chosen subjectively. In this paper, we build on these combining studies in two important ways.

Firstly, given that there are many different VaR and ES forecasting methods available, we consider the combination of forecasts from large pools of forecasts. We consider pools consisting of up to 90 different methods. With a pool as large as this, our work can be framed as aiming to extract the wisdom from the predictions of a crowd of experts (see \citealp{galton1907}).

The second way in which our work extends the existing literature is that the availability of a large number of individual forecasts leads us to propose new combining methods. In addition to the use of the mean and median as simple robust benchmarks, we also consider an estimator of the mode based on kernel density estimation. For the performance-based weighted combinations of \cite{taylor2020forecast}, to avoid overfitting in the estimation of the large number of combining weights, we incorporate regularisation, which has become popular in machine learning. Like \cite{wang2022forecasting}, we consider the trimmed mean, but we introduce a wider set of trimming methods, which draw on interval forecast combining methods from the decision analysis literature (see, e.g., \citealp{park2015aggregating}). To enable this, we essentially view the forecasts of VaR and ES from each method as the bounds of an interval forecast. We also adapt the interval forecast combining approaches of \citet{taylor2025depth}, which are based on statistical depth, and \cite{gaba2017combining}, which infers a probability distribution from each VaR and ES forecast pair, and then performs probability averaging. These new approaches lead to a sizeable number of potential combining methods. Given the fundamental proposal to combine when faced with multiple forecasts, it seems natural, when we have forecasts from a sizeable number of combining methods, to combine those forecasts \citep{gunter1989n}. We do this via simple averaging and an extension of the smooth transition combining of \cite{deutsch1994combination}. In all combining methods, we optimise parameters using one of the joint scores of \cite{Fissler2016}.

In Section \ref{preliminaries_sec}, after describing the structure of our empirical study and joint scoring functions for VaR and ES, we list the 90 individual methods that we use to produce the VaR and ES forecasts. Section \ref{sec:combining_sec} presents the combining methods, including a review of previously proposed methods, as well as our new methods. Section \ref{sec:empirical} describes our empirical study, which compares the combining methods for pools consisting of between 90 and just six methods. Section \ref{sec:summary} provides a summary and concluding remarks.


\section{Preliminaries}\label{preliminaries_sec}

\subsection{Structure of our empirical forecasting study}\label{Data_sec}

We used daily high, low and closing prices, as well as high-frequency 5-minute realised variance (RV), for the following six market indices: AORD, DAX, FTSE 100,  Hang Seng, Nikkei and S\&P 500. We calculated daily log returns using daily closing prices, and daily ranges as the difference between the daily high and low log prices. The values of returns and ranges are scaled by 100 to express them as a percentage. The 5-minute RV for each day is defined as the sum of the 5-minute squared returns on that day. The square root of RV is the realised volatility. The dataset corresponding to each index consisted of 5400 daily observations ending on 19 May 2022. The start date differed across the indices due to different public holidays in each country. Descriptive statistics of the daily returns are presented in Table \ref{t:data_summary}.

\begin{table}[ht!]
\captionsetup{font=small}
\caption{\label{t:data_summary} Descriptive statistics of daily log returns across the six markets.}
\vspace{-10pt}
\small
\renewcommand{\arraystretch}{1.2}
\begin{center}
\begin{tabular}{llllllll}
\toprule
Market&Mean&Median&Min&Max&Std&Skewness&Kurtosis\\ \midrule
AORD&0.0155&0.0691&-7.261&4.533&0.946&-0.730&9.118\\
DAX&0.0134&0.0763&-11.863&12.027&1.456&-0.165&9.637\\
FTSE 100&0.0024&0.0343&-10.137&9.485&1.162&-0.308&10.471\\
Hang Seng&0.0048&0.0369&-13.582&13.407&1.449&-0.018&10.618\\
Nikkei&0.0054&0.0438&-12.111&13.235&1.481&-0.381&9.140\\
S\&P 500&0.0186&0.0636&-12.670&10.642&1.233&-0.405&13.829\\
\bottomrule
\end{tabular}
\end{center}
\end{table}
For each individual method in the pool of methods to be combined, we used a rolling window of 1800 days to re-estimate method parameters each day, and hence produce out-of-sample day-ahead VaR and ES forecasts for the final 3600 days. This period of 3600 days was the focus of our study of forecast combining. For each combining method, we also used a rolling window of 1800 days to re-estimate parameters each day, and produce out-of-sample combined forecasts of VaR and ES for the final 1800 days. In Section \ref{sec:empirical}, we also consider the performance of the combining methods when parameter estimation uses notably shorter rolling windows.

We focus on the VaR and ES for probability level $\alpha$ =2.5$\%$, given its widespread consideration in the literature. We write the return, VaR and ES in period \textit{t} as  $r_t$, $VaR_t$ and $ES_t$. With a pool of $M$ individual forecasting methods, at each forecast origin, we produced $M$ forecasts of the 2.5\% VaR and ES, which we express as $\widehat{VaR}_{m,t}$ and $\widehat{ES}_{m,t}$, for $m=1,2,\ldots,M$. In Section \ref{sec:empirical}, we present results for several pool sizes, ranging from $M=6$ to $90$.

As we define VaR as a conditional quantile in the lower tail of the returns distribution, the VaR and ES are both negative, and the ES must exceed the VaR in the sense that the ES will be a value below the VaR. Therefore, for each combining method $m$, it is clearly necessary that  $\widehat{ES}_{m,t}$ $\leq$  $\widehat{VaR}_{m,t}$. This is analogous to the condition that two quantile estimates do not cross. To ensure there is no crossing, in some of the methods that we consider, alongside combinations of VaR forecasts, we combine forecasts of the $spacing$ between VaR and ES, which we write for method $m$ as $\Delta_{m,t}= {VaR}_{m,t}-{ES}_{m,t}$.

\subsection{Joint score for VaR and ES}\label{joint_scoring_sec}

\cite{Fissler2016} provide the joint scoring function in  (\ref{e:joint-score}) for the VaR and ES.
\begin{align} \label{e:joint-score}
S(VaR_t,ES_t,r_t) = & \quad (I(r_t \le VaR_t) - \alpha) {G_1} ({VaR_t}) - I(r_t \le VaR_t) {G_1} ({r_t}) \notag \\ & + {G_2} ({ES_t}) ({ES_t} - {VaR_t} + I(r_t \le VaR_t) ({VaR_t} - {r_t}) / \alpha ) \notag \\ & - \zeta_2 ({ES_t}) + a({r_t}).
\end{align}
$I$ is the indicator function; ${G_1}$ is an increasing function; ${\zeta_2}$ is an increasing and convex function; ${G_2}$=${\zeta'_2}$; and $a$ is a function that does not depend on the VaR or ES. If ${\zeta_2}$ is strictly increasing and strictly concave, the scoring function is strictly consistent.

To estimate combining method parameters in Section \ref{sec:combining_sec} and several of the individual forecasting methods in Section \ref{individual_method_sec}, we use the asymmetric Laplace (AL) score of \cite{taylor2019al}, who shows that it is a member of the class of scores in  (\ref{e:joint-score}), and that it links to quantile regression. For this score, ${G_1}=0$,  ${G_2}=-1/x$,  ${\zeta_2}(x)=-\ln(-x)$ and $a=1-\ln(1-\alpha)$. We also use the AL joint score, along with several other joint scores, to evaluate VaR and ES out-of-sample forecast accuracy in Section \ref{sec:empirical}.

\subsection{Individual methods for VaR and ES forecasting}\label{individual_method_sec}

We consider 90 individual methods for VaR and ES forecasting, which we list in Table \ref{t:ind_methods_detail} in Appendix \ref{sec:90_methods_appendix}. The methods reflect a wide spectrum of modelling and data utilisation strategies, drawing on information from daily returns, high-low price ranges, and realized volatility measures. This heterogeneity in both methodology and data inputs ensures a rich and varied pool of forecasts, capturing different aspects of return dynamics and tail risk, making them well-suited for forecast combining, which thrives on diversity \citep{gerlach2017semi}.

\begin{figure}[ht!]
     \centering
\includegraphics[width=\textwidth]{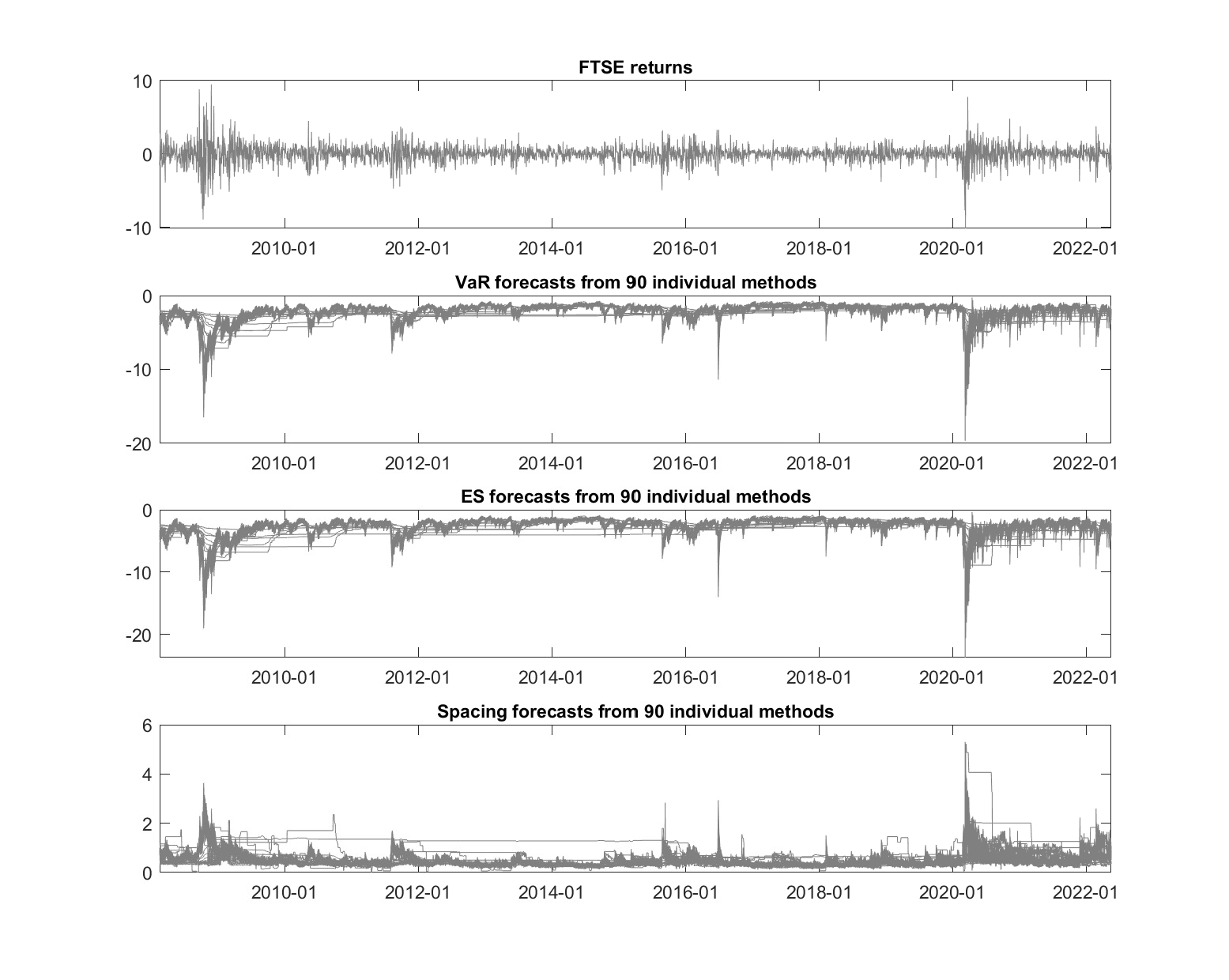}
\vspace{-40pt}
\caption{\label{Fig:VaR_ES_All} For the 3600-day period ending on 19 May 2022, the FTSE 100 returns and forecasts of the VaR, ES and spacing produced by the 90 individual forecasting methods.}
\end{figure}

Figure \ref{Fig:VaR_ES_All} presents the 3600 out-of-sample VaR and ES forecast  produced by the 90 individual methods for the FTSE 100 data, along with a plot of the returns. The figure also presents a plot of the spacings between the VaR and ES forecasts, as defined at the end of Section \ref{Data_sec}. The figure shows quite substantial variety in the forecasts from the 90 methods. In the remainder of this subsection, we provide brief descriptions of the methods.

\subsubsection{Simplistic methods}\label{simple_method_sec}

\noindent \textbf{Historical Simulation (HS) (individual methods 1-4).}
This nonparametric method employs the sample quantile as the VaR forecast, and for the ES forecast, uses the sample average of returns exceeding the VaR forecast. We included the following three common choices for the sample size: 100, 250 and 500 days. For example, for a sample size of 100, the VaR and ES were estimated from the 100 days up to and including the forecast origin. We also implemented the method using 1800 days, which corresponded to the in-sample period for each forecast origin.

\noindent \textbf{Gaussian (individual methods 5-8).}
This is a parametric approach that uses a Gaussian distribution with zero mean and sample standard deviation of the 100, 250, 500 or 1800 days up to and including the forecast origin.

\noindent \textbf{EWMA (individual method 9).} This method forecasts the variance using an exponentially weighted moving average with smoothing parameter set as the widely-used value of 0.94. A Gaussian distribution is then used to produce the VaR and ES forecasts.

\subsubsection{GARCH-based models}\label{garch_method_sec}

\noindent \textbf{GARCH (individual methods 10-18).} We used the GARCH(1,1) model with the following three different distributions in the maximum likelihood estimation: Gaussian, Student $t$, and the skewed $t$ of \cite{hansen1994autoregressive}. For each estimation approach, we employed three different procedures to produce VaR and ES forecasts based on the GARCH variance forecasts. The first procedure used the same distribution that was used in the maximum likelihood estimation. The second procedure was the method of \citep{mcneil2000estimation}, which applies extreme value theory to the standardised residuals. The third procedure used filtered historical simulation, which involves the application of historical simulation to the 1800 standardised residuals.

\noindent \textbf{GARCH with Range (individual methods 19-27).} These methods are similar to individual methods 10-18, with the key difference that the squared residual in the GARCH(1,1) model is replaced by the square of the daily high-low price range  (\citealp{molnar2016high}). The range is defined as $R_t= log(h_t)-log(l_t)$, with $h_t$ and $l_t$ representing the highest and the lowest price on day $t$.

\noindent \textbf{GARCH with RV (individual methods 28-36).} These methods are the same as individual methods 10-18, except that realised variance is used instead of the squared residual in the GARCH(1,1) model \citep{engle2002new}.

\noindent \textbf{GJR-GARCH (individual methods 37-63).} These 27 methods are the same as individual methods 10-36, except that the GARCH(1,1) model is replaced by the GJR-GARCH(1,1) model of \cite{glosten1993relation}. This model incorporates the leverage effect, which has been shown to be beneficial (see, for example, \citealp{ning2015volatility}). When the squared range and RV are employed in GJR-GARCH, the return is still used within the asymmetric component of the model to enable modelling of the leverage effect.

\subsubsection{CAViaR-based models}\label{caviar_method_sec}

\noindent \textbf{CAViaR (individual methods 64-69).} These methods use the AL joint score to estimate joint models for the VaR and ES, with a CAViaR model specified for the VaR and two alternative formulations for the ES \citep{taylor2019al}. The first ES formulation simply sets the ES as a constant multiple of the VaR, while the second specifies the ES as the sum of the VaR and an autoregressive model for the spacing between the ES and the VaR. In Table \ref{t:ind_methods_detail} in Appendix \ref{sec:90_methods_appendix}, we refer to these two formulations as $Multiplicative$ and $Additive$, respectively. In terms of CAViaR models, we used the symmetric absolute value (SAV), asymmetric slope (AS), and indirect-GARCH (IG).

\noindent \textbf{CAViaR-Range (individual methods 70-75).} These are the same as individual methods 64-69, except that the range $R_t$ is used instead of the return in the CAViaR model \citep{gerlach2020semi}. The one exception to this is in the asymmetric component of the CAViaR-AS model, where the return is still used to enable the leverage effect to be modelled.

\noindent \textbf{CAViaR-RV (individual methods 76-81).} These are the same as individual methods 70-75, except that realised volatility is used instead of range in the CAViaR models \citep{gerlach2020semi}.

\subsubsection{CARE-based models}\label{caviar_method_sec}

\noindent \textbf{CARE (individual methods 82-84).} \cite{tayl2008}  notes that the expectation of observations exceeding an expectile can be written as a simple function of the expectile, and uses this as the basis for VaR and ES estimation. The VaR at probability level $\alpha$ is approximated by the expectile for which the proportion of observations exceeding the expectile is  $\alpha$. To model the expectile, \cite{tayl2008}  draws on the CAViaR models of \cite{caviar}  to introduce SAV, AS and IG conditional autoregressive expectile (CARE) models.

\noindent \textbf{CARE-Range (individual methods 85-87).} These are the same as individual methods 82-84, except that range is used instead of the return in the CARE model \citep{gerlach2017semi}, with the one exception being the use of the return within the asymmetric component of the CARE-AS model to enable modelling of the leverage effect.

\noindent \textbf{CARE-RV (individual methods 88-90).} These are the same as individual methods 85-87, except that the CARE models use realised volatility instead of range \citep{gerlach2017semi}.

\section{Combining methods}\label{sec:combining_sec}

We now present combining methods for VaR and ES forecasts. Section  \ref{sec:aggregation_no_his_sec}  consists of 12 methods that do not rely on the historical accuracy of the individual methods to be combined. Section \ref{sec:performance_agg_sec} presents five performance-based weighted combining methods. Section \ref{sec:comb_agg_sec} describes two approaches for combining the forecasts produced by combining methods. Table 2 summarises the 19 combining methods. Even though a significant interest in this paper is combining a large pool of forecasts, the table's penultimate column notes that many of the methods we consider are not limited to this situation. As indicated in the final column of the table, many of the methods have not previously been considered for VaR and ES. Some have been used in other contexts and require little adaptation for VaR and ES. However, this is not the case for others, as will be apparent in Sections \ref{sec:aggregation_no_his_sec} to \ref{sec:comb_agg_sec}.

\begin{table}[ht!]
\captionsetup{font=small}

\caption{\label{t:methods_detail} Summary of the VaR and ES forecast combining methods.}
\vspace{-10pt}
\small
\renewcommand{\arraystretch}{1.2}
\begin{center}
\begin{tabular}{lccc}
\toprule
Name&\multicolumn{1}{p{2.7cm}}{\centering Historical  \\ \vspace{-3pt} accuracy needed?} &\multicolumn{1}{p{2.5cm}}{\centering Sizeable \\  \vspace{-3pt} pool needed?}   &\multicolumn{1}{p{2.5cm}}{\centering New \\  \vspace{-3pt} method?}   \\ \midrule
\multicolumn{4}{l}{\textit{Methods based on measures of central tendency (Section \ref{sec:aggregation_mean})}}\\
 \textit{  }  Simple average& No& No&No\\
\textit{  } Median&No&Yes&Yes\\
\textit{  } Mode&No&Yes&Yes\\ \arrayrulecolor{lightgray}\hline
 \multicolumn{4}{l}{\textit{Methods based on interval forecast combining (Section \ref{trimmed_mean_sec})}}\\
\textit{  } Symmetric trimmed mean&No&Yes&No\\
\textit{  } Exterior trimmed mean&No&Yes&Yes\\
\textit{  } Interior trimmed mean&No&Yes&Yes\\
\textit{  } Lower trimmed mean&No&Yes&Yes\\
\textit{  } Higher trimmed mean&No&Yes&Yes\\
\textit{  } Flexible trimmed mean&No&Yes&Yes\\
\textit{  } Probability average&No&No&Yes\\
 \textit{  } Halfspace deepest&No&No&Yes \\
 \textit{  } Simplicial  deepest& No&No&Yes \\
\arrayrulecolor{lightgray}\hline
 \multicolumn{4}{l}{\textit{Performance-based weighted combining (Section \ref{sec:performance_agg_sec})}}\\
\textit{  } Relative score&Yes&No&No\\
\textit{  } Relative score with weighted median&Yes&No&Yes\\
\textit{  } Minimum score&Yes&No&No\\
\textit{  } Minimum score with ratio&Yes&No&Yes\\
\textit{  } Minimum score with regularisation&Yes&No&Yes\\ \arrayrulecolor{lightgray}\hline
 \multicolumn{4}{l}{\textit{Combinations of forecasts produced by combining methods (Section \ref{sec:comb_agg_sec})}}\\
\textit{  } Smooth transition combining&Yes&No&Yes\\
\textit{  } Mean of all combinations&Yes&No&Yes\\

\arrayrulecolor{black}\bottomrule
\end{tabular}

\end{center}

\end{table}

\subsection{Combining not based on historical performance of individual methods}\label{sec:aggregation_no_his_sec}

In this section, we consider combining methods that do not require a record of historical accuracy for the individual methods. An advantage of such combining methods is that they do not involve the estimation of performance-based weights, which can be challenging for a large pool of individual methods, or when a comparable record of past forecast accuracy is not available from all methods.

\subsubsection{Combining based on statistical measures of central tendency}\label{sec:aggregation_mean}

\noindent \textbf{Simple average.} For each forecast origin, the simple average of all $M$ VaR forecasts and simple average of all $M$ ES forecasts are used as the combined forecasts. Such averaging is a standard benchmark in studies of forecast combining.

\noindent \textbf{Median.} For each forecast origin, the median of the $M$ VaR forecasts and the median of the $M$ ES forecasts are used as the combined forecasts. Median combining has the appeal of robustness to outlying forecasts. Although often considered in other forecast combining contexts, we are not aware of its previous use for VaR and ES, which is understandable because the few papers in this area have not considered a large pool of forecasts.

\begin{figure}[htp]
     \centering
\includegraphics[width=\textwidth]{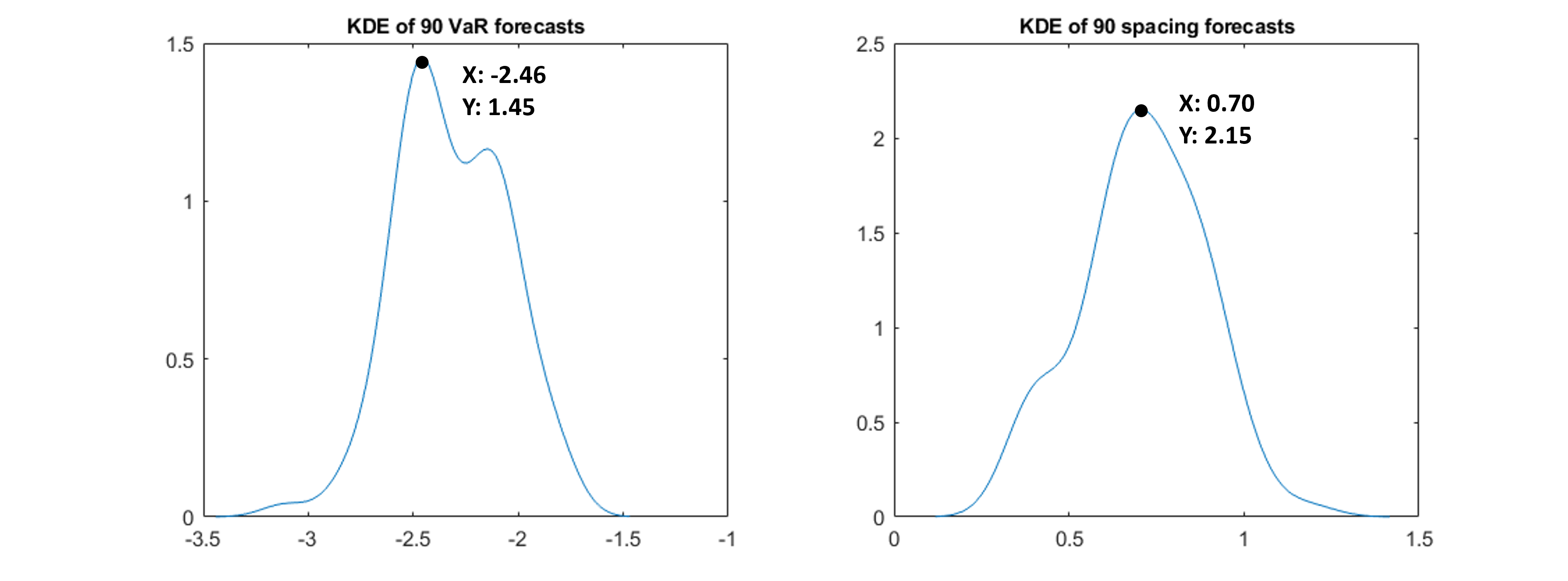}
\vspace{-20pt}
\caption{\label{Fig:kde} Mode-based combining applied to the pool of 90 VaR and spacing forecasts for the FTSE 100 at forecast origin 18 May 2022.}
\end{figure}

\noindent \textbf{Mode.} We follow  \cite{kourentzes2014neural} who apply kernel density estimation (KDE) to a set of forecasts, and then note the mode of the resulting density. While their focus is point forecasting, we use the approach for VaR and ES. To be more precise, we apply KDE to the $M$ VaR forecasts to obtain their mode, and to ensure that this VaR forecast is exceeded by the resulting ES forecast, we obtain the mode of the $M$ spacings, also by applying KDE. We optimised the KDE bandwidth separately for the VaR and spacing. We illustrate the approach in Figure \ref{Fig:kde} for the pool of size $M=90$, consisting of all individual methods described in Section \ref{individual_method_sec}. The figure shows the densities produced by KDE for the FTSE 100 at the final forecast origin used in our study. In each plot, the $x$-coordinate of the mode is the combined forecast.

\subsubsection{Combining based on methods for interval forecast combining}\label{trimmed_mean_sec}

The combining methods that we propose in this section are inspired by interval forecast combining methods from the decision analysis literature. To draw on this literature, we treat each ES and VaR forecast pair as, respectively, the lower and upper bounds of an interval forecast.

\begin{figure}[htp]
     \centering
\includegraphics[width=0.9\textwidth]{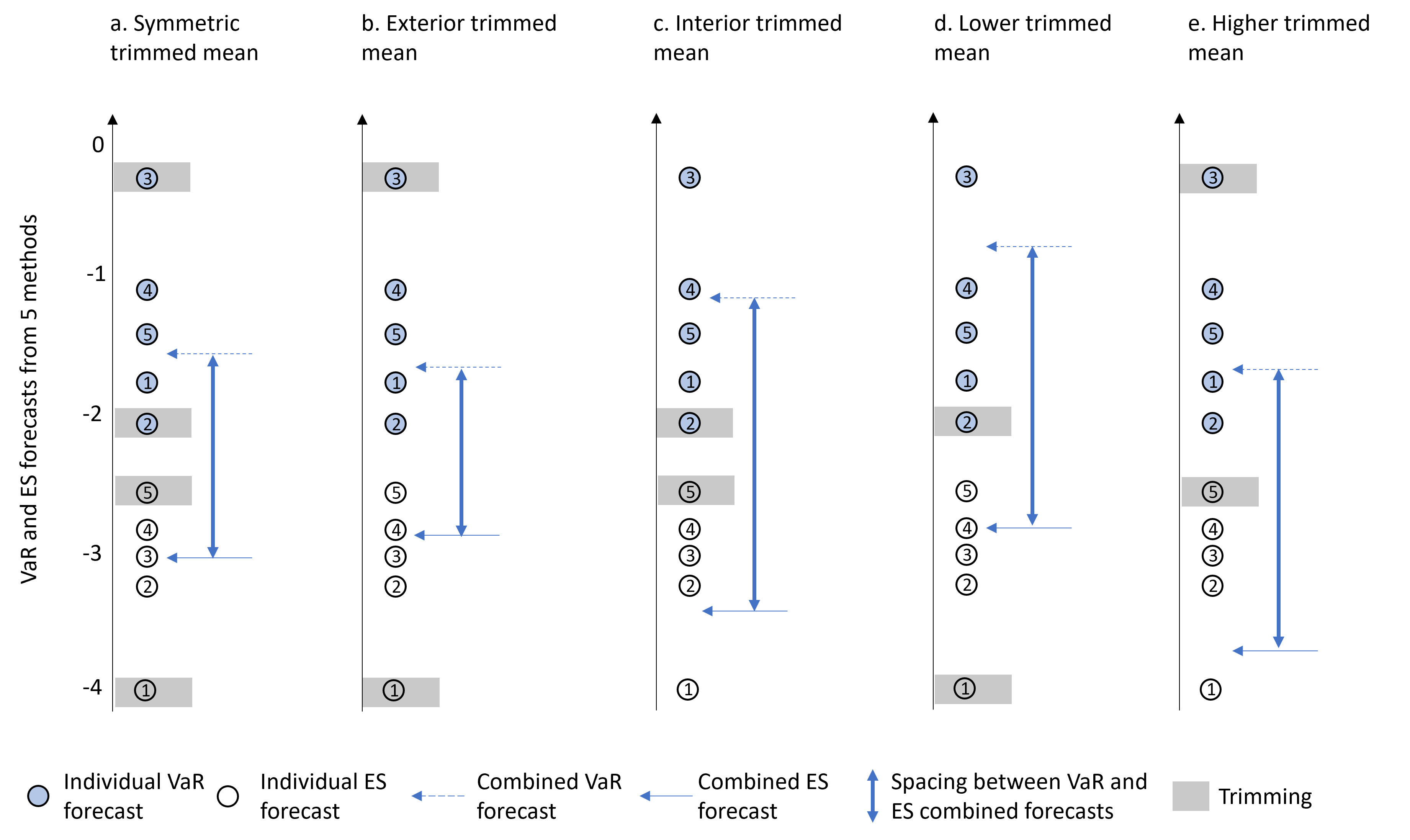}
\vspace{-10pt}
\caption{\label{Fig:trimming} Illustration of trimmed mean combining for VaR and ES forecasts from five individual methods. }
\end{figure}

\noindent \textbf{Symmetric trimmed mean.} This is based on the interval forecast combining method of \cite{park2015aggregating}, and involves the standard trimmed mean, which is motivated by robust estimation. We apply this separately for VaR and ES. For the VaR, we average the forecasts remaining after trimming the $N$ lowest-valued and $N$ highest-valued forecasts, where $N$ is the trimming parameter, which is an integer between 0 and $M/2-1$. The same approach is applied to the ES forecasts. An illustration is provided in Figure \ref{Fig:trimming}.a for an example with five VaR and ES forecasts. We optimised the trimming parameter, which we chose to be the same for the VaR and ES in this method. This trimmed mean method was used by \cite{wang2022forecasting}, who subjectively set the trimming parameter as $N=5$ when combining predictions from $M=15$ individual forecasting methods.

\noindent \textbf{Exterior trimmed mean.} This draws on the interval forecast combining method of \cite{gaba2017combining}, which enables the width of the combined interval forecast to be reduced. In our implementation, the method controls the spacing between the combined forecasts of the VaR and ES. More specifically, we average the VaR forecasts remaining after trimming the $N$ highest-valued VaR forecasts, and average the ES forecasts remaining after removing the $N$ lowest-valued ES forecasts, where $N$ is a trimming parameter, which is an integer between 0 and $M-1$. Figure \ref{Fig:trimming}.b provides an illustration of the method.

\noindent \textbf{Interior trimmed mean.} \cite{gaba2017combining} also proposed a trimmed mean approach that enables the widening of the combined interval forecast. In our adaptation of this for VaR and ES, we average the VaR forecasts remaining after trimming the $N$ lowest-valued VaR forecasts, and average the ES forecasts remaining after removing the $N$ highest-valued ES forecasts, where $N$ is an integer between 0 and $M-1$. The method is illustrated in Figure \ref{Fig:trimming}.c.

\noindent \textbf{Lower trimmed mean.} This method is not an adaptation of an existing interval forecast combining method. The method aims to address the situation where the VaR and ES forecasts are generally too low. We average the VaR forecasts after trimming the $N$ lowest of these forecasts, and we average the ES forecasts remaining after removing the $N$ lowest of these forecasts. $N$ is an integer between 0 and $M-1$. The method is illustrated in Figure \ref{Fig:trimming}.d.

\noindent \textbf{Higher trimmed mean.} This method is also not an adaptation of an existing interval forecast combining method. The method is suitable when the VaR and ES forecasts are generally too high. We remove the $N$ highest-valued ES forecasts, and the $N$ highest-valued VaR forecasts. The remaining VaR and ES forecasts are then averaged. $N$ is an integer between 0 and $M-1$. Figure \ref{Fig:trimming}.e illustrates the approach.

\begin{figure}[ht!]
     \centering
\includegraphics[width=\textwidth]{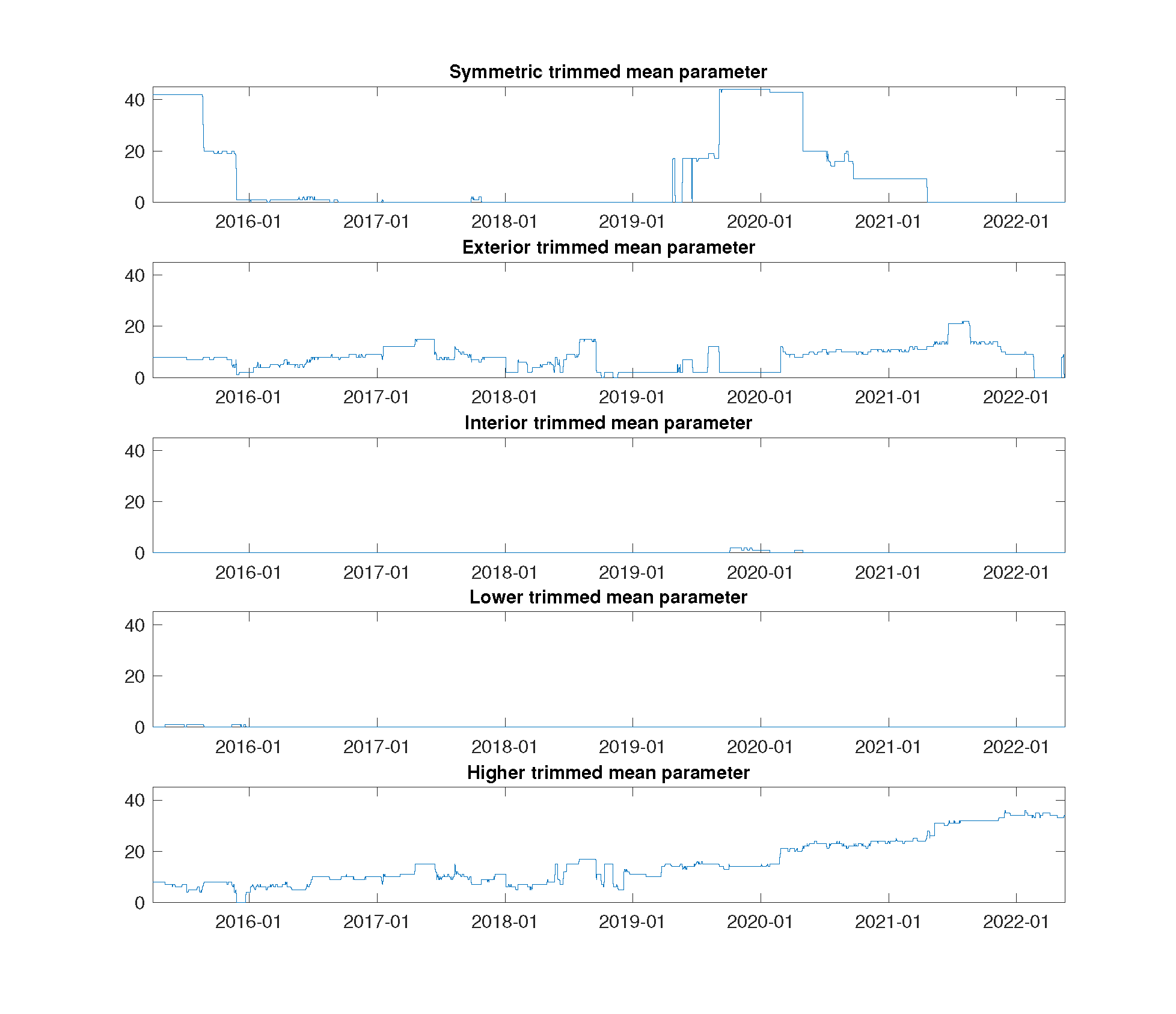}
\vspace{-50pt}
\caption{\label{Fig:trimming_parameter} The trimming parameter $N$ optimised for each forecast origin for the trimmed mean combining methods applied to the pool of 90 individual VaR and ES forecasts for the FTSE 100.}
\end{figure}

For the trimmed mean methods discussed so far, Figure \ref{Fig:trimming_parameter} presents the trimming parameter optimised at the 1800 forecast origins for the FTSE 100, with the pool consisting of all $90$ individual methods described in Section \ref{individual_method_sec}. For symmetric trimming, there are periods with sizeable trimming, including the first year of the pandemic. The trimming parameter is mostly non-zero for the exterior trimmed mean, suggesting the VaR and ES forecasts were generally too high and low, respectively. For the interior trimmed mean and lower trimmed mean, as there is negligible trimming, these methods closely match the simple average. For the higher trimmed mean, the trimming parameter values suggest that the VaR and ES forecasts were generally insufficiently extreme, with this being increasingly the case over time.

\vspace{-12pt}
\vspace{-5pt}
\noindent \textbf{Flexible trimmed mean.} This method encompasses the five trimmed mean methods described so far. It has a separate trimming parameter for the VaR and ES. Each parameter can be a positive or negative integer. A positive value indicates the number of forecasts to be trimmed from the lower end of the range of forecasts, while a negative value is the number to be trimmed from the upper end of the range.

\begin{figure}[htp]
     \centering
\includegraphics[width=\textwidth]{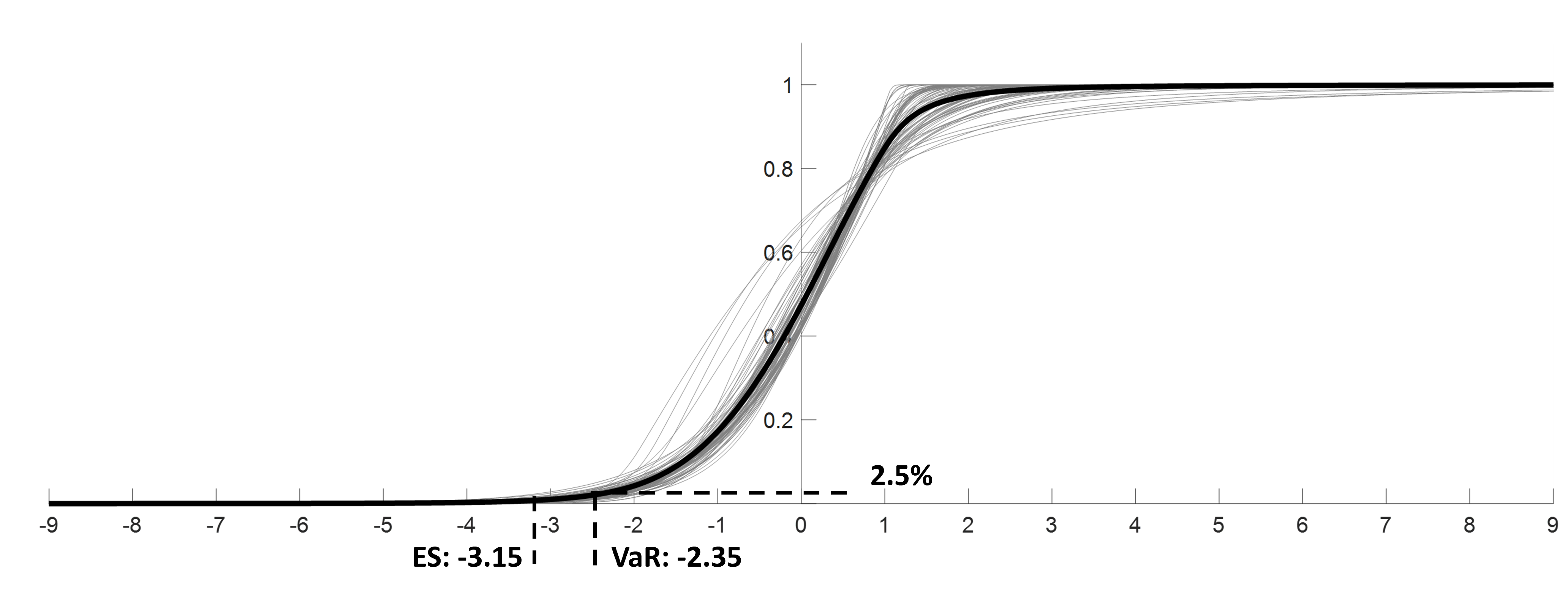}
\caption{\label{Fig:prob_avg_90} For the FTSE 100 at forecast origin 18 May 2022, the 90 CDFs corresponding to the VaR and ES forecasts of the pool of 90 individual methods. The black CDF is the result of probability averaging. The VaR and ES of the black CDF are shown, and these are the resulting combined forecasts.}
\end{figure}

\noindent \textbf{Probability averaging.} The probability averaging approach to interval forecast combining, proposed by \cite{gaba2017combining} draws on the work of \cite{lichtendahl2013better}, who show how, when averaging forecasts of a cumulative distribution function (CDF), the CDF resulting from averaging probabilities will have greater spread than the CDF produced by averaging quantiles. Separately averaging the forecasts of each bound of an interval amounts to averaging quantiles. As an alternative to this interval forecast averaging, the probability averaging approach of \cite{gaba2017combining} first views each individual method's interval forecast as coming from a Gaussian distribution. Probability averaging is then applied to the set of Gaussian CDFs, with the resulting CDF providing the combined interval forecast. It follows from the findings of \cite{lichtendahl2013better} that the resulting combined interval forecast will be wider than the interval resulting from simply averaging the lower bounds and averaging the upper bounds.

To adapt this approach for VaR and ES, we first obtain a CDF corresponding to the VaR forecast and ES forecast from each individual method. Probability averaging is then applied to the pool of CDFs, with the resulting CDF providing the VaR and ES, which are used as the combined forecasts. The results of \cite{lichtendahl2013better} imply that this approach will lead to more extreme VaR and ES forecasts than those provided by the simple average method, which averages the VaR forecasts and averages the ES forecasts. Obtaining the CDF from an individual method is straightforward if it is parametric because the VaR and ES forecasts are derived from a CDF forecast. Table \ref{t:ind_methods_detail} in Appendix \ref{sec:90_methods_appendix} shows that 23 of the 90 individual methods in our study were of this type. For the other 67 methods, we must infer a CDF from the VaR forecast and ES forecast. To infer a CDF, we assume it is skewed $t$ of the type introduced by \cite{hansen1994autoregressive}. We use this distribution because of its flexibility, with parameters for the standard deviation $\sigma$, skewness $\lambda$, and degrees of freedom $\nu$. We employ a grid search over these three parameters to find the specific skewed $t$ distribution for which the VaR and ES are closest to the VaR and ES forecasts in terms of Euclidean distance. If this leads to more than one solution, we select the CDF with lowest skewness. We provide additional details of our implementation of the probability averaging method in  Appendix \ref{sec:prob_averaging_implementation}.

For the pool consisting of all 90 individual methods, Figure~\ref{Fig:prob_avg_90} shows the 90 CDFs corresponding to the VaR and ES forecasts of the 90 methods for the final forecast origin of the FTSE 100. Also shown is the CDF resulting from the probability averaging of these CDFs, and the VaR and ES of this CDF. These VaR and ES values are the resulting combined forecasts.

\noindent \textbf{Halfspace deepest.} For interval forecast combining, \citet{taylor2025depth} suggests that a potential weakness of median combining is that the median of each interval bound is obtained separately. The issue is that if one bound of an individual method's interval forecast is an outlier, it perhaps reduces the appeal of using the other bound. To try to capture the available information better, \citet{taylor2025depth} treats each interval forecast as a bivariate point, and uses statistical depth to identify the deepest point. Drawing on this, we treat the $M$ pairs of individual VaR and ES forecasts as $M$ bivariate points, and use halfspace depth \citep{tukey1975mathematics} to find the deepest. The VaR and ES corresponding to this deepest point are the combined forecasts. For a given point $p$, the halfspace depth is the minimum proportion of observations lying in any closed halfspace whose boundary passes through $p$.

\begin{figure}[htp]
     \centering
\includegraphics[width=.7\textwidth]{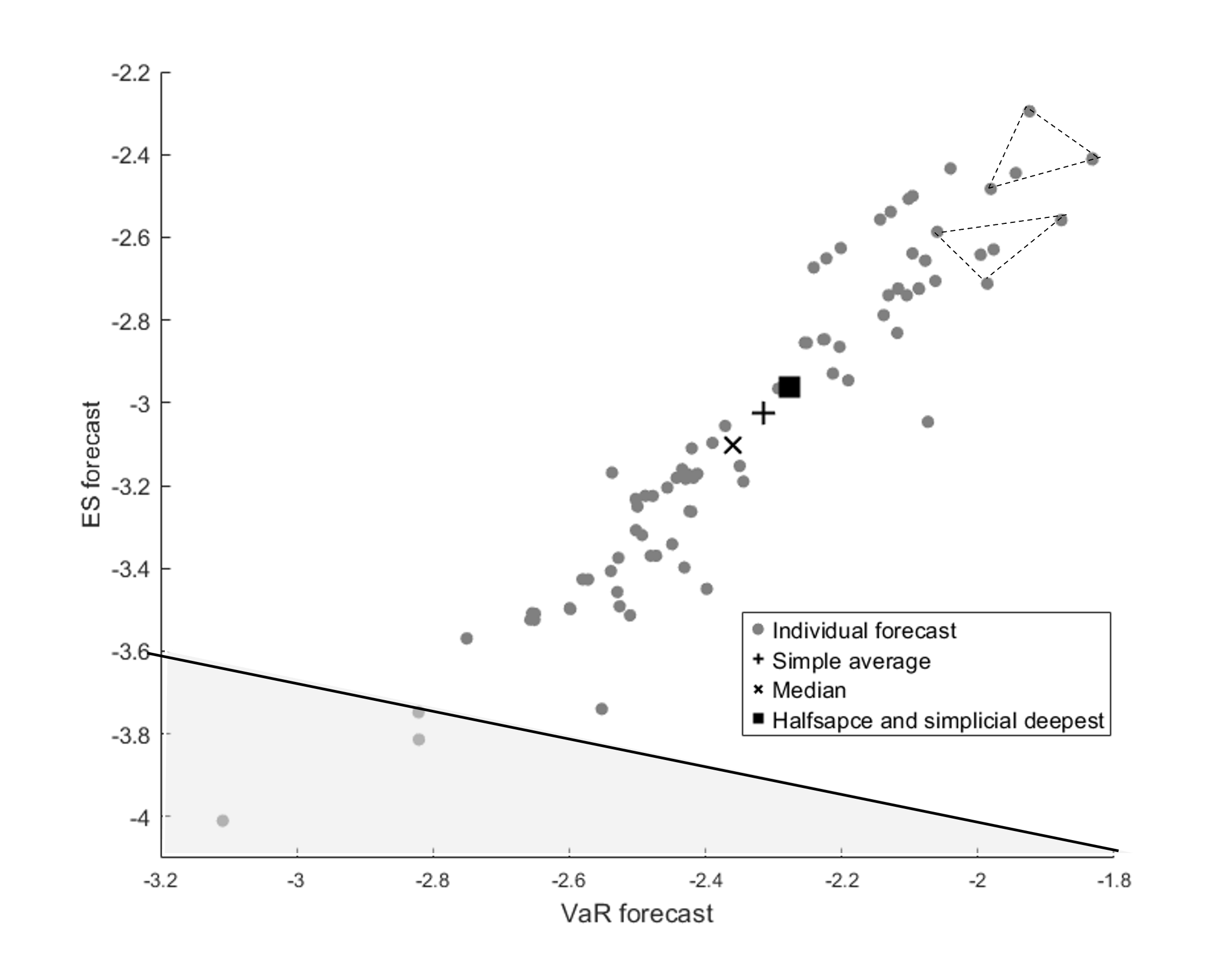}
\caption{\label{Fig:depth} For the FTSE 100 at forecast origin 18 May 2022, points representing the VaR and ES forecasts of the pool of 90 individual methods and the following combining methods: simple average, median and deepest point. The grey region is an example of a halfspace bounded by a line through a selected point. The triangles are examples of simplices used when computing simplicial depth.}
\end{figure}

For the pool consisting of all 90 individual methods, and for the final forecast origin of the FTSE 100, Figure \ref{Fig:depth} displays as points the pairs of VaR and ES forecasts from the 90 methods and the halfspace deepest point. The VaR and ES for this point is the combined forecast. The shaded region illustrates a halfspace whose boundary passes through a selected point. The figure also shows the points corresponding to simple average combining and median combining, and it is interesting to see that these differ from the halfspace deepest point. We also note that, by contrast with the simple average and median combining, halfspace depth delivers a pair of VaR and ES forecasts from one individual method, which in the example of Figure \ref{Fig:depth} is CARE-Range-AS. This can be viewed, informally, as the median forecasting method.

\noindent \textbf{Simplicial deepest}. We follow \cite{taylor2025depth} in also applying the simplicial depth of \cite{liu1990notion} to assess bivariate depth. For a point $p$, simplicial depth is defined as the proportion of closed simplices, formed from three points in the data cloud, that contain $p$. The simplices are triangles for bivariate data, and in Figure \ref{Fig:depth}, we provide examples of triangles formed with three of the points as vertices. For the set of 90 points in Figure \ref{Fig:depth}, the deepest point found using halfspace and simplicial depth was the same.

\subsection{Performance-based weighted combining methods}\label{sec:performance_agg_sec}

The combining methods that we consider in this section use the historical accuracy of the individual methods to produce performance-based weighted combinations.

\noindent \textbf{Relative score.} This is one of two methods introduced by  \cite{taylor2020forecast}. It involves combining weights that are inversely proportional to one of the joint scoring functions in (\ref{e:joint-score}) summed over the estimation sample. The method has only one parameter $\lambda$, and uses the same simplex set of weights to combine the VaR and ES forecasts. It is presented in (\ref{relative_comb}).
\begin{eqnarray}\label{relative_comb}
   \widehat{VaR}_{c,t}& = & \sum_{m=1}^{M} w_m \widehat{VaR}_{m,t}, \\ \nonumber
   \widehat{ES}_{c,t}& = &\sum_{m=1}^{M} w_m \widehat{ES}_{m,t}, \\ \nonumber
    \omega_m&=& \frac{\exp \left(  -\lambda \sum_{j=1}^{t-1} S \left( \widehat{VaR}_{m,j},\widehat{ES}_{m,j}, r_j  \right) \right)} {\sum_{m=1}^{M} \exp \left(  -\lambda \sum_{j=1}^{t-1} S \left( \widehat{VaR}_{m,j},\widehat{ES}_{m,j}, r_j  \right) \right) },
\end{eqnarray}
where $w_m$ is the weight for method $m$. If $\lambda$ is close to zero, the method becomes the simple average, while if $\lambda$ is large, the individual method with the best historical accuracy is selected.

\begin{figure}[ht!]
\includegraphics[width=1.3\textwidth,center]{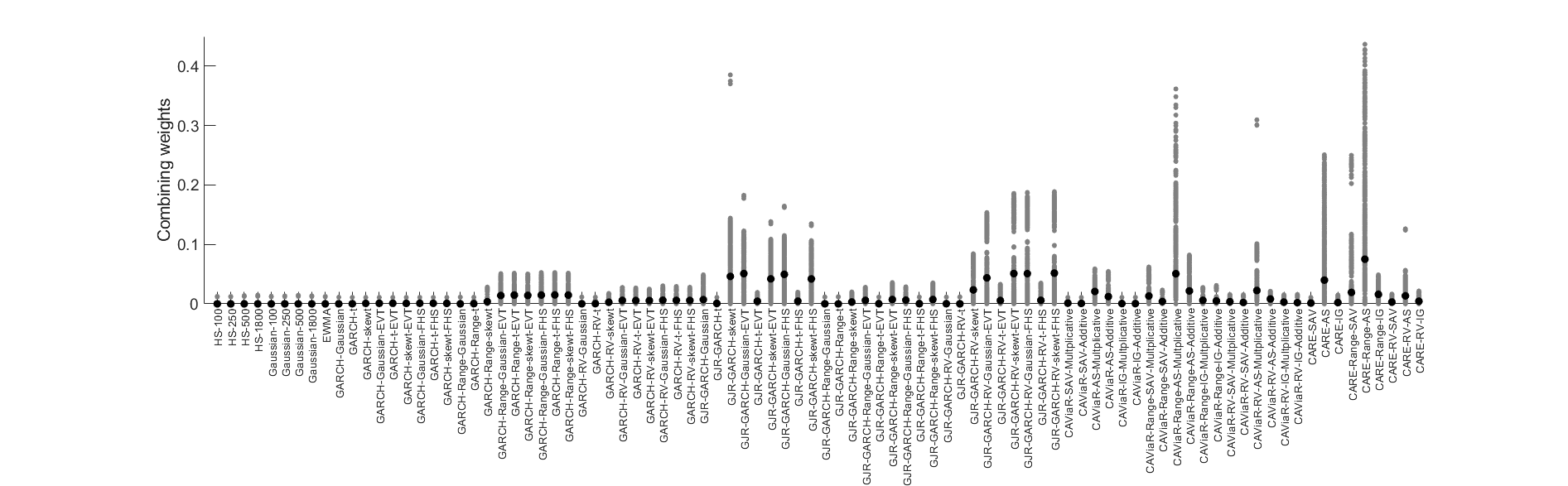}
\vspace{-10pt}
\caption{\label{Fig:relative_score} For the 1800 forecast origins for the FTSE 100, the relative score combining weights optimised for the pool of 90 individual methods. The black points show the average of the combining weights for each individual method in the pool.}
\end{figure}

As with the parameters for all combining methods, we optimised $\lambda$ afresh for each of the 1800 forecast origins. For the pool consisting of all 90 individual methods, the resulting 1800 values of the combining weight for each of the 90 methods are presented in Figure \ref{Fig:relative_score}. The figure shows that the simplistic methods, such as historical simulation, received low weights. In terms of the GARCH models, it is interesting to see the largest weights for GJR-GARCH, particularly with realised variance. By contrast, using range, rather than realised variance, led to the highest weights for the CAViaR-based models and CARE models.

\noindent \textbf{Relative score with weighted median.} Motivated by the appeal of robust estimation, we propose a weighted median approach using the same formulation for the weight $w_m$ presented for relative score combining in (\ref{relative_comb}).

\noindent \textbf{Minimum score.} This is the second combining method introduced by \cite{taylor2020forecast}. It is motivated by relative score combining being potentially overly-restrictive in using the same combining weights for the VaR and ES. To address this, while also ensuring the combined ES forecast exceeds the combined VaR forecast, separate combining weights, $w_m^{VaR}$ and $w_m^\Delta$, are obtained for the VaR and for the spacing between VaR and ES. The method is presented in (\ref{mini_comb_spacing}).
\begin{eqnarray}\label{mini_comb_spacing}
   \widehat{VaR}_{c,t}& = & \sum_{m=1}^{M} w_m^{VaR} \widehat{VaR}_{m,t}, \\
   \hat{\Delta}_{m,t}& = & \widehat{VaR}_{m,t} -\widehat{ES}_{m,t}, \nonumber \\
   \widehat{ES}_{c,t}& = &\widehat{VaR}_{c,t} - \sum_{m=1}^{M} w_m^\Delta \hat{\Delta}_{m,t}, \nonumber \\
   s.t. &\quad&  \boldsymbol{\omega}^{VaR} \in [0,1], \sum \boldsymbol{\omega}^{VaR} =1 \nonumber \\
        &\quad& \boldsymbol{\omega}^{\Delta} \in [0,1], \sum \boldsymbol{\omega}^{\Delta} =1.  \nonumber
\end{eqnarray}

\begin{figure}[h!]
     \centering
\includegraphics[width=1.3\textwidth,center]{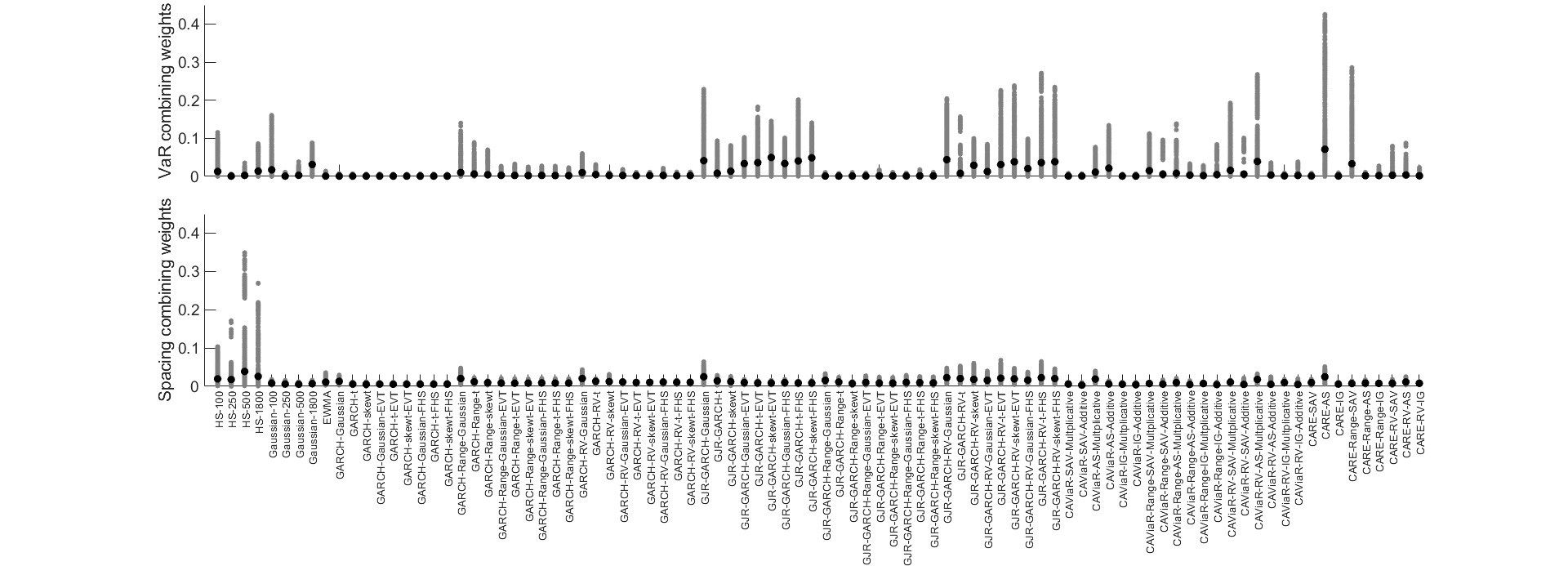}
\vspace{-10pt}
\caption{\label{Fig:minimum_score} For the 1800 forecast origins for the FTSE 100, the minimum score VaR and spacing combining weights, optimised for the pool of 90 individual methods. The black points show the average of the combining weights for each individual method in the pool.}
\end{figure}

With separate simplex sets of combining weights for the VaR and spacing, there are $2(M-1)$ weights to be estimated. For the pool consisting of all 90 individual methods, we have $M=90$, which implies 178 weights. The optimised values of these weights for the 1800 forecast origins for the FTSE 100 are shown in Figure \ref{Fig:minimum_score}.  The two plots show that higher weights resulted for the simplistic methods than shown in Figure \ref{Fig:relative_score} for relative score combining. Furthermore, it is interesting to see the particularly high weights in the bottom panel in Figure \ref{Fig:minimum_score}, suggesting that the simplistic methods have more to offer in combinations of the spacing forecasts than in combinations of VaR forecasts.

\noindent \textbf{Minimum score with ratio.} The CAViaR-based formulations of \cite{taylor2019al} model the ES as either the sum or product of the VaR and another term. Given that minimum score combining models the ES as the sum of the VaR and the spacing, the work of \cite{taylor2019al} prompts a new version of this method, where the ES is modelled as the product of the combined forecast for the VaR and the combined forecast for the ratio of the ES to the VaR, as in (\ref{mini_comb_ratio}).
\begin{eqnarray}\label{mini_comb_ratio}
   \widehat{VaR}_{c,t}& = & \sum_{m=1}^{M} w_m^{VaR} \widehat{VaR}_{m,t}, \\
   \hat{R}_{m,t}& = & \frac{\widehat{ES}_{m,t}}{\widehat{VaR}_{m,t}},  \nonumber \\
   \widehat{ES}_{c,t}& = &\widehat{VaR}_{c,t}  \sum_{m=1}^{M} w_m^R \hat{R}_{m,t}, \nonumber \\
   s.t. &\quad&  \boldsymbol{\omega}^{VaR} \in [0,1], \sum \boldsymbol{\omega}^{VaR} =1 \nonumber  \\
        &\quad& \boldsymbol{\omega}^{R} \in [0,1], \sum \boldsymbol{\omega}^{R} =1. \nonumber
\end{eqnarray}




\noindent \textbf{Minimum score with regularisation.} Recognising that minimum score combining in (\ref{mini_comb_spacing}) involves a potentially large number of combining weights, we propose regularised estimation using the loss function in (\ref{ridge_obj}), which incorporates ridge penalty terms in the joint score in (\ref{e:joint-score}).
\begin{eqnarray}\label{ridge_obj}
    \text{Loss} \left( {\boldsymbol{\omega}^{VaR}, \boldsymbol{\omega}^\Delta} \right)&=&\left(  \frac{1}{T} \sum_{t=h}^{T+h-1} \textit{S} \left( \widehat{VaR}_{c,t}, \widehat{ES}_{c,t}, r_t \right) \right)\\ \nonumber
     && + \lambda_1 \sum_{m=1}^{M} \left( \omega_{m}^{VaR} \right)^2 + \lambda_2  \sum_{m=1}^{M}  \left( \omega_{m}^\Delta \right)^2  \\ \nonumber
     && s.t. \quad  \boldsymbol{\omega}^{VaR} \in [0,1], \boldsymbol{\omega}^{\Delta} \in [0,1], \sum \omega_{m}^{VaR}=1, \sum \omega_{m}^\Delta=1,
\end{eqnarray}
where $h$ represents the first period in the window of $T$ observations used for estimation. In our empirical study, we selected the regularization parameters $\lambda_1$ and $\lambda_2$ using holdout validation with a 75:25 split of the in-sample data into training and validation subsets.

Due to the simplex constraints, the ridge penalty terms are minimised when the weights are equal. Therefore, the regularisation shrinks the weights towards equality, rather than towards zero. This was recognised by \cite{diebold2023aggregation} in a study of density forecast combining. Like them, we feel this is potentially useful because it amounts to shrinkage towards the simple average, which is a natural benchmark in forecast combining. Note that Lasso penalties would have no impact on weight estimation in the presence of the simplex constraints.

\citet[Chapter 3]{happersberger2021advancing} also considers regularised estimation for combinations of VaR and ES forecasts. However, his formulation differs from ours in that he applies regularisation to unconstrained linear combinations of VaR forecasts and ES forecasts, which contrasts with our inclusion of simplex constraints, and our focus on VaR and spacing forecasts, in order to avoid the possibility of the ES exceeding the VaR.

\begin{figure}[h!]
     \centering
\includegraphics[width=1.3\textwidth,center]{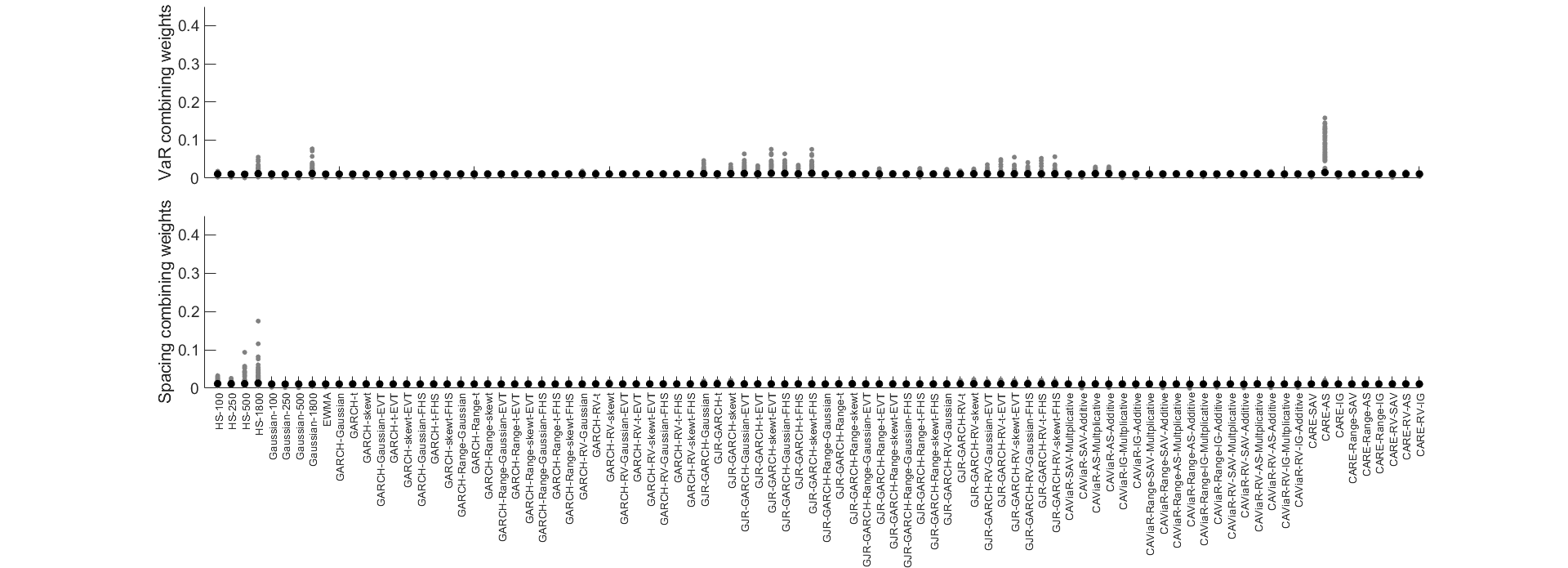}
\vspace{-20pt}
\caption{\label{Fig:ridge_weights} For the 1800 forecast origins for the FTSE 100, the VaR and spacing combining weights for minimum score combining with regularisation, optimised for the pool of 90 individual methods. The black points show the average of the combining weights for each individual method in the pool.}
\end{figure}

For the pool consisting of all 90 individual methods, Figure \ref{Fig:ridge_weights} presents the combining weights resulting from minimum score with regularisation for the 1800 forecast origins of the FTSE 100 series. The impact of the regularisation towards equal weights can be seen by noting that the weights in this figure show much less variation among the methods than in Figure \ref{Fig:minimum_score} for minimum score with no regularisation.

\subsection{Combinations of forecasts produced by combining methods}\label{sec:comb_agg_sec}

\noindent \textbf{Smooth transition combining (STC).} \cite{deutsch1994combination} propose a method that provides smooth transition between two combinations of point forecasts. We adapt the method here to enable smooth transition between the forecasts from two VaR and ES combining methods. Given the VaR forecasts, $\widehat{VaR}_{t,1}$ and $\widehat{VaR}_{t,2}$, ES forecasts, $\widehat{ES}_{t,1}$ and $\widehat{ES}_{t,2}$, and spacings, $\widehat{\Delta}_{t,1}$ and $\widehat{\Delta}_{t,2}$ , from two methods, we produce combined forecasts as follows:
\begin{eqnarray}\label{e:stc_model}
    \widehat{VaR}_{t,stc} &=& F\left(z_t;\ubeta\right) \widehat{VaR}_{t,1} + \left(1-F\left(z_t;\ubeta\right) \right) \widehat{VaR}_{t,2},\nonumber \\ \nonumber
    \widehat{ES}_{t,stc} &=& \widehat{VaR}_{t,stc} - \left( F\left(z_t;\ubeta\right)  \widehat{\Delta}_{t,1} +  (1-F\left(z_t;\ubeta\right))  \widehat{\Delta}_{t,2} )\right),\\ \nonumber
    F\left(z_t;\ubeta\right)&=& \left(  1+\text{exp} \left(  - \left( \beta_0 + \beta_1 z_{t}  \right) \right)  \right)^{-1}, \nonumber
\end{eqnarray}
where $F\left(z_t;\ubeta\right)$ is a combining weight that has the form of a transition function, $z_t$ is a variable that controls the transition between the forecasts of the two methods, and $\beta_0$ and $\beta_1$ are parameters estimated using the in-sample data.

We specify the first method as the simple average and the second as relative score combining. We chose these methods because they are examples of a simplistic combining method and a performance-based weighted combining method. In terms of the transition variable $z_t$, we surmised that the relative performance of these two methods depends on the level of volatility in the data. To capture this, we chose $z_t$ to be the average of the VaR forecasts from the 90 individual methods, i.e.,  $\widehat{VaR}_{t,1}$. Other transition variables could also be considered, such as intraday risk measures, option-implied volatility, or a variable reflecting the state of the economy, as in the macroeconomic point forecasting application of \cite{deutsch1994combination}.

\begin{figure}[htp]
     \centering
\includegraphics[width=1.2\textwidth,center]{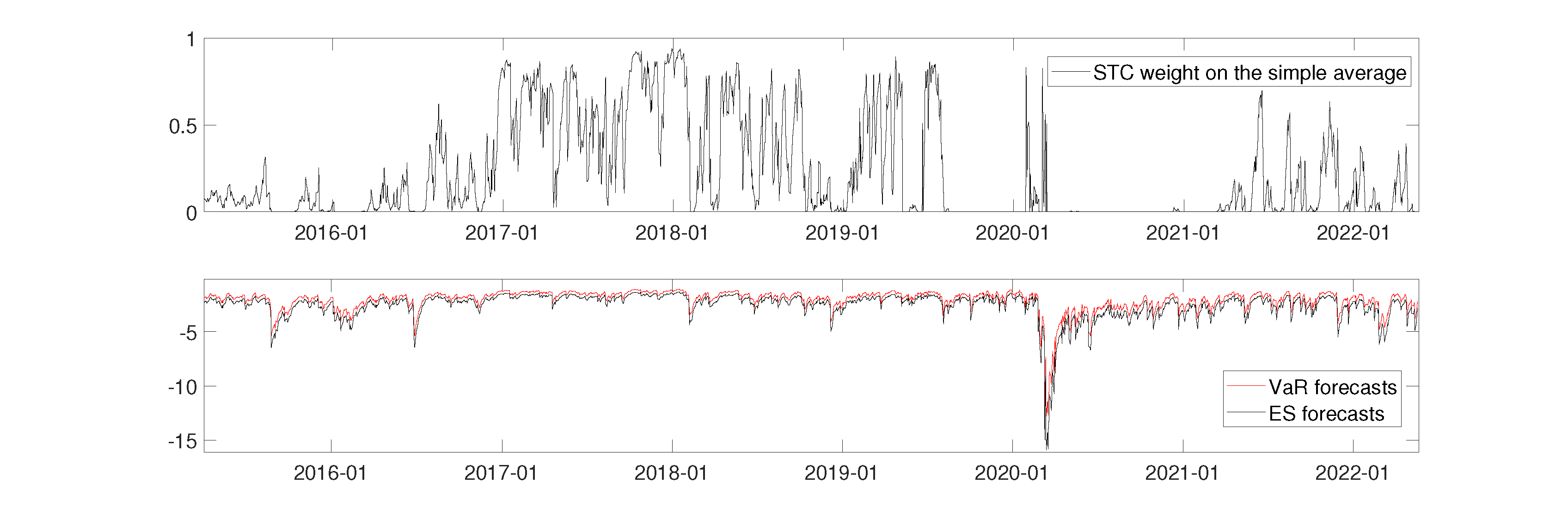}
\vspace{-20pt}
\caption{\label{Fig:stc_weights} For the out-of-sample period for the FTSE 100 and the pool of 90 individual methods, the smooth transition combining weight on the simple average when combined with relative score combining, and the resulting VaR and ES forecasts.}
\end{figure}

For the pool consisting of all 90 individual methods and the out-of-sample evaluation period (observations $3601$ to $5400$) of the FTSE 100, the function $F\left(z_t;\ubeta\right)$ is plotted in the upper panel of Figure \ref{Fig:stc_weights}. This function is the weight on the simple average forecasts. The figure's lower panel provides a plot of the VaR and ES forecasts resulting from smooth transition combining. Assuming that the fluctuations in the VaR and ES forecasts are a reasonable reflection of the time-varying volatility, the figure shows that the combining weight on the simple average forecasts tended to be lower when the volatility was high.

\noindent \textbf{Simple average of combining methods.} In Section \ref{individual_method_sec}, we built up a sizeable set of diverse and competing forecasting methods. We then took the view that their forecasts should be combined. As we now have a sizeable set of diverse and competing forecast combining methods, it seems only natural to combine their forecasts, as suggested by \cite{gunter1989n} in a point forecasting context. We implemented this by calculating the simple average of the forecasts from the other 18 forecast combining methods in this section.

\section{Empirical results}\label{sec:empirical}

In this section, we compare the out-of-sample performance of the 19 combining methods described in Section \ref{sec:combining_sec}. As a benchmark, we also report the results for a method we have called \textit{dynamic selection}, which produces forecasts at each origin using the individual method (out of the $M$ methods in the pool) for which the AL joint score was lowest for the most recent in-sample moving window. Following standard practice in the literature, we evaluated the out-of-sample VaR and ES forecasts using calibration tests and scoring functions, which we describe in Section \ref{calib_scores_sec}. In Section \ref{evaluation_large_pool_sec}, we report the results for the large pool of $M=90$ VaR and ES forecasts. Section \ref{reduced_pool_sec} evaluates whether better results can be achieved using smaller diverse pools. For Sections \ref{evaluation_large_pool_sec} and \ref{reduced_pool_sec}, all individual and combining methods were implemented using rolling window estimation samples of 1800 days. In Section \ref{reduced_sample_sec}, we investigate whether better results can be produced with a smaller sample size.

\subsection{Calibration tests and scoring functions}\label{calib_scores_sec}

For probability level 2.5\%, if the proportion of out-of-sample VaR forecasts exceeded by the corresponding return is equal to 2.5\%, the VaR forecasts are said to be \textit{unconditionally} calibrated. If, in addition, the exceedances occur randomly over time, the VaR forecasts are \textit{conditionally} calibrated. We implemented the unconditional calibration test of \cite{Kupiec1995} (UC), and for conditional calibration, we used the tests of \cite{chri1998} (CC) and the dynamic quantile test of \cite{caviar} (DQ) using four lags. The ES forecasts are calibrated if the conditional mean of the VaR exceedances matches the mean of the ES forecasts. To test this, we employed the bootstrap test of \cite{mcneil2000estimation}, which, for the VaR exceedances, tests whether the average of the standardised difference between the return and the ES forecast is equal to zero. We standardised by dividing the difference by the VaR forecast.

We also evaluated the VaR and ES forecasts using the same set of five scoring functions employed by \cite{taylor2020forecast}, averaging each over the 1800-day out-of-sample period. The first of the five scoring functions is the \textit{quantile score} (QS). This is the loss function used within quantile regression, and it is the only one of the five scores that evaluates just VaR forecast accuracy. The second scoring function is the AL joint score, which we also used for estimating combining method parameters. The remaining three are also versions of the joint scoring function in (\ref{e:joint-score}). They are the versions proposed by \cite{nolde2017} (NZ), \cite{Fissler2016} (FZG), and \cite{acesze2014} (AS). The QS can also be viewed as a special case of the joint score in (\ref{e:joint-score}), enabling us to conveniently summarise the five scores in Table \ref{t:scoring_functions}, by specifying the corresponding form of each of the functions in (\ref{e:joint-score}).

\begin{table}[ht!]
    \centering
    \small
    \caption{Functions used within the joint VaR and ES scoring function of (\ref{e:joint-score}) to give the five different versions of the score used in our study: QS, AL, NZ, FZG, and AS.}
    \label{t:scoring_functions}
    \begin{tabular}{@{}lcccc@{}}
        \toprule
        & $G_1(x)$ & $G_2(x)$ & $\zeta_2(x)$ & $a(x)$ \\
        \midrule
        QS& $x$ & 0& 0 & $\alpha x$ \\
        AL & 0 & $-1/x$ & $-\ln(-x)$ & $1 - \ln(1 - \alpha)$ \\
        NZ & 0 & $1/2(-x)^{-1/2}$ & $-(-x)^{1/2}$ & 0 \\
        FZG & $x$ & $\exp(x)/(1 + \exp(x))$ & $\ln(1 + \exp(x))$ & $\ln(2)$ \\
        AS & $-1/2Wx^2$ & $\alpha x$ & $1/2\alpha x^2$ & 0 \\
        \bottomrule
    \end{tabular}
\end{table}

For each of the six indices, we calculated the skill score for each method, which is the percentage by which the average score for that method was better than a benchmark method. We chose historical simulation based on 250 days (HS-250) as the benchmark, as this is often used as a simple benchmark in studies of VaR and ES. To summarise the skill scores across the six indices, we calculated the geometric mean of the ratios of the average score of each method to the average score of HS-250, then subtracted this from 1 and multiplied the result by 100.

\subsection{Results for the pool of 90 methods}\label{evaluation_large_pool_sec}

For the pool of $M=90$ methods, let us first consider calibration results. In Table \ref{t:calibration_1800}, for each combining method, we report the number of indices, out of the six, for which calibration was rejected at the significance level of 5\%. Lower values are preferable in this table. The only methods for which calibration could not be rejected for all four tests and all six indices are probability averaging, minimum score and minimum score with ratio.

\setlength{\tabcolsep}{8pt}
\begin{table}[ht!]
\captionsetup{font=small}
\centering
\small
\renewcommand{\arraystretch}{1.2}
\caption{Calibration test results for combining applied to the pool of $M=90$ individual
methods. The number of indices for which calibration was rejected at the 5$\%$ significance level using the UC, CC, DQ and ES tests. 1800 days used for estimation of individual methods and combining methods.\\}
	\label{t:calibration_1800}
	\centering
    \vspace{-10pt}
	\begin{tabular}{lccccc}
    \toprule
Method&UC&CC&DQ&ES\\ \midrule
\multicolumn{5}{l}{\textit{Benchmark method}} &\\
\textit{  } Dynamic selection bench&1&1&1&3\\
\arrayrulecolor{lightgray}\hline
\multicolumn{5}{l}{\textit{Methods based on measures of central tendency}} &\\
\textit{  } Simple average&0&2&1&0\\
\textit{  } Median&0&1&1&0\\
\textit{  } Mode&0&1&0&0\\
\arrayrulecolor{lightgray}\hline
  \multicolumn{5}{l}{\textit{Methods based on interval forecast combining}} &\\
\textit{  } Symmetric trimmed mean&0&1&1&0\\
\textit{  } Exterior trimmed mean&0&2&1&2\\
\textit{  } Interior trimmed mean&1&1&2&0\\
\textit{  } Lower trimmed mean&0&1&1&2\\
\textit{  } Higher trimmed mean&0&1&1&0\\
\textit{  } Flexible trimmed mean&1&1&1&1\\
\textit{  } Probability average&0&0&0&0\\
\textit{  } Halfspace deepest&0&1&1&0\\
\textit{  } Simplicial  deepest&0&1&1&0\\
 \arrayrulecolor{lightgray}\hline
  \multicolumn{5}{l}{\textit{Performance-based weighted combining}} &\\
\textit{  } Relative score&0&2&1&0\\
\textit{  } Relative score with weighted median&0&2&2&1\\
\textit{  } Minimum score&0&0&0&0\\
\textit{  } Minimum score with ratio&0&0&0&0\\
\textit{  } Minimum score with regularisation&0&1&1&0\\
\arrayrulecolor{lightgray}\hline
  \multicolumn{5}{l}{\textit{Combinations of forecasts produced by combining methods}} &\\
\textit{  } Smooth transition combining&0&1&2&0\\
\textit{  } Simple average of all combinations&0&1&1&0\\

\arrayrulecolor{black}\bottomrule
\end{tabular}
\end{table}

For the scoring functions, skill scores are presented in the left half of Table \ref{t:Skill_Rank_1800}. Higher values are preferable, with the best three methods for each score highlighted with grey shading. To enable an additional summary of performance across the six indices, we calculated each method’s rank for each score and index, and then averaged these ranks across the six indices to give the results in the right half of Table \ref{t:Skill_Rank_1800}. Lower values are preferable, with shading again used to highlight the three best methods for each score.

The first point to note in Table \ref{t:Skill_Rank_1800} is that, for all five scores, all the combining methods outperform the dynamic selection method at the top of Table \ref{t:Skill_Rank_1800}. This is a sophisticated benchmark method, so the results are encouraging for forecast combining. In terms of which form of combining performed the best, the shading in Table \ref{t:Skill_Rank_1800} highlights notably strong results for the higher trimmed mean, probability averaging, the three minimum score methods, and the simple average of all combinations. It is interesting that these methods come from different sections within the table, providing some justification for our consideration of methods based on interval forecasting, performance-based weighting, and combinations of combined forecasts.

The success of the higher trimmed mean shows that the large pool of individual forecasts tended to be insufficiently extreme, as trimming the higher values and then averaging led to improved results over the simple average method, and indeed very competitive results overall. We had anticipated that the flexible trimmed mean would be the best performing of the trimming approaches because its flexibility is such that it encompasses all the other forms of trimmed mean. However, the results indicate that the flexibility, achieved via the use of two parameters, has come at a cost to accuracy. Given that the higher trimmed mean has performed well, it is not surprising that the results of probability averaging are also very competitive, as this method also produces VaR and ES forecasts that are more extreme than those from the simple average method. Among the other methods based on interval forecast combining, Table \ref{t:Skill_Rank_1800} shows that simplicial and halfspace depth, which are bivariate analogies of the median, did not match the performance of median combining, which involves the simpler univariate median.

 For the performance-based weighted combining methods, the slight superiority of the minimum score methods over relative score matches the finding of \cite{taylor2020forecast} for a small pool of individual forecasts and estimation for the 1\% probability level. Table \ref{t:Skill_Rank_1800} shows that there was little benefit in extending the relative score and minimum score methods to the new versions, although regularisation did lead to improved average ranks for the minimum score method.

We found the results in the bottom row of the table interesting. They show that when faced with a choice between a diverse set of combining methods, competitive results can be produced by taking the average of the forecasts produced by all of them.

\setlength{\tabcolsep}{5pt}
\begin{table}[ht!]
\captionsetup{font=small}
\centering
\small
\renewcommand{\arraystretch}{1.2}
	\caption{Scoring function results for combining methods applied to the pool of $M=90$ individual methods. For each score, skill scores and ranks averaged across the six stock indices. Higher skill scores and lower rank values are better, with grey shading highlighting the three best performing methods in each column. 1800 days used for estimation of individual methods and combining methods. \\}
	\label{t:Skill_Rank_1800}
	\centering
    \vspace{-10pt}
	\begin{tabular}{lccccc|ccccc}\toprule
\multicolumn{1}{c}{} & \multicolumn{5}{c}{Skill score} \vline & \multicolumn{5}{c}{Rank} \\		
Method&QS&AL&NZ&FZG&AS&QS&AL&NZ&FZG&AS\\  \midrule
\multicolumn{5}{l}{\textit{Benchmark method}} &\\
\textit{  } Dynamic selection&15.4&9.7&9.4&7.2&24.6&   14.7&15.2&15.0&15.5&14.0
 \\
\arrayrulecolor{lightgray}\hline
\multicolumn{5}{l}{\textit{Methods based on measures of central tendency}} &\\
\textit{  } Simple average&16.0&10.2&9.8&7.6&25.6& 9.7&9.0&9.2&8.3&9.8
\\
\textit{  } Median&16.0&10.2&9.8&7.6&25.9&8.8&9.7&9.3&9.7&9.5

\\
\textit{  } Mode&15.8&10.0&9.6&7.4&25.5&13.0&13.5&12.7&13.2&13.0
\\
\arrayrulecolor{lightgray}\hline
  \multicolumn{5}{l}{\textit{Methods based on interval forecast combining}} &\\
\textit{  } Symmetric trimmed mean&15.9&10.1&9.7&7.5&25.6& 11.8&11.5&11.8&11.5&12.2
\\
\textit{  } Exterior trimmed mean&16.0&10.2&9.8&7.6&25.6&9.5&10.8&10.3&11.0&10.3
\\
\textit{  } Interior trimmed mean&15.6&10.1&9.7&7.6&25.3& 13.3&11.3&12.3&8.8&11.7
\\
\textit{  } Lower trimmed mean&15.7&10.0&9.6&7.4&25.2&12.7&13.5&13.2&13.0&12.7
\\
\textit{  } Higher trimmed mean& 16.2 & \cg{10.4} & 9.9 & 7.7 & 26.0 & \cg{7.0}&\cg{6.8}&\cg{6.2}&\cg{6.8}&\cg{7.5}
\\
\textit{  } Flexible trimmed mean&15.5&9.7&9.5&7.2&25.5& 11.5&10.8&11.0&9.7&11.3
\\
\textit{  } Probability average&  16.2 &	 \cg{10.4} &	 9.9 &	 \cg{7.8} &	 26.2 &\cg{6.7}&\cg{5.5}&\cg{5.7}&\cg{6.0}&\cg{6.7}
\\
\textit{  } Halfspace deepest&15.8&10.1&9.7&7.5&25.5& 12.3&12.8&12.3&13.3&12.8
\\
\textit{  } Simplicial  deepest&15.9&10.1&9.7&7.5&25.5& 11.8&11.8&12.3&12.0&12.8
\\
 \arrayrulecolor{lightgray}\hline
  \multicolumn{5}{l}{\textit{Performance-based weighted combining}} &\\
\textit{  } Relative score&\cg{16.3}&10.3& 9.9 &7.7&26.2&9.8&11.7&10.5&12.8&8.8
\\
\textit{  } Relative score with weighted median&16.1&10.1&9.8&7.5&25.6&11.5&14.0&13.2&15.2&11.3
\\
\textit{  } Minimum score&\cg{16.4}&\cg{10.4}&\cg{10.0}&\cg{7.8}&\cg{26.5}& 9.3&8.3&9.5&9.0&8.3
\\
\textit{  } Minimum score with ratio&\cg{16.3}&\cg{10.4}&\cg{10.0}& \cg{7.8} &\cg{26.3}&9.8&9.7&9.7&9.0&10.2 \\
\textit{  } Minimum score with regularisation&16.1&10.3& 9.9 &7.7&25.9& \cg{7.8} & \cg{7.2} & 7.8 &\cg{6.0}&8.8

\\
\arrayrulecolor{lightgray}\hline
  \multicolumn{5}{l}{\textit{Combinations of forecasts produced by combining methods}} &\\
\textit{  } Smooth transition combining&16.2&10.3&9.9&7.7&26.1& 10.3&10.5&10.7&12.3&10.0
\\
\textit{  } Simple average of all combinations&\cg{16.3}&\cg{10.5}&\cg{10.0}&\cg{7.8}&\cg{26.3}& 8.5&6.3 & \cg{7.3} & 6.8&\cg{8.2}
\\
 \arrayrulecolor{black}\bottomrule

\end{tabular}
\end{table}

The model confidence set (MCS) testing of \cite{hansen2011_MCS} identifies a group of models that, with a specified probability, includes the best model with respect to a chosen scoring function. Models not contained in the MCS are viewed as less likely to be the best. Using a 75\% confidence level, we applied MCS testing separately for each of the five scores, employing the equivalence test based on the Diebold-Mariano test, and the one-sided elimination rule referred to as $T_{max,M}$ by \cite{hansen2011_MCS}. In Table \ref{t:MCS_1800}, for each of the five scores, we report the number of stock indices for which each method is included in the MCS. Higher counts in this table are preferable. The table confirms the finding from Table \ref{t:Skill_Rank_1800} that the poorest results were produced by the dynamic selection benchmark method.

\setlength{\tabcolsep}{5pt}
\begin{table}[ht!]
\captionsetup{font=small}
\centering
\small
\renewcommand{\arraystretch}{1.2}
	\caption{Model confidence set results for combining methods applied to the pool of $M=90$ individual methods. For each score, number of stock indices (out of the six) for which each method is included in the model confidence set for a 75\% confidence level. Higher values are better, with grey shading highlighting the methods that are included in the model confidence set for all six indices. 1800 days used for estimation of individual methods and combining methods.\\}
	\label{t:MCS_1800}
	\centering
    \vspace{-10pt}
	\begin{tabular}{lccccccc}\toprule
	
Method&QS&AL&NZ&FZG&AS\\  \midrule
\multicolumn{5}{l}{\textit{Benchmark method}} &\\
\textit{  } Dynamic selection&4&2&4&4&4\\
\textit{  } Simple average&\cg{6}&5&5&5&\cg{6}\\
\textit{  } Median&\cg{6}&5&5&5&\cg{6}\\
\textit{  } Mode&5&4&5&5&\cg{6}\\
\arrayrulecolor{lightgray}\hline
\multicolumn{5}{l}{\textit{Methods based on measures of central tendency}} &\\
\textit{  } Symmetric trimmed mean&\cg{6}&5&5&5&\cg{6}\\
\textit{  } Exterior trimmed mean&5&4&5&5&\cg{6}\\
\textit{  } Interior trimmed mean&5&4&5&5&\cg{6}\\
\textit{  } Lower trimmed mean&5&4&5&5&\cg{6}\\
\textit{  } Higher trimmed mean&4&5&5&5&5\\
\textit{  } Flexible trimmed mean&5&4&5&5&5\\
\textit{  } Probability average&4&5&5&5&5\\
\textit{  } Halfspace deepest&5&4&5&5&\cg{6}\\
\textit{  } Simplicial  deepest&\cg{6}&5&5&5&\cg{6}\\
 \arrayrulecolor{lightgray}\hline
  \multicolumn{5}{l}{\textit{Performance-based weighted combining}} &\\
\textit{  } Relative score&5&\cg{6}&5&5&5\\
\textit{  } Relative score with weighted median&4&4&4&4&4\\
\textit{  } Minimum score&\cg{6}&5&5&5&\cg{6}\\
\textit{  } Minimum score with ratio&\cg{6}&5&4&4&\cg{6}\\
\textit{  } Minimum score with regularisation&\cg{6}&5&5&5&\cg{6}\\
\arrayrulecolor{lightgray}\hline
  \multicolumn{5}{l}{\textit{Combinations of forecasts produced by combining methods}} &\\
\textit{  } Smooth transition combining&5&5&5&5&5\\
\textit{  } Simple average of all combinations&\cg{6}&\cg{6}&5&5&\cg{6}\\
\arrayrulecolor{black}\bottomrule

\end{tabular}
\end{table}

\subsection{Reduced pool size }\label{reduced_pool_sec}

In this section, we compare the combining methods in terms of their performance for different pool sizes. In addition to the large pool of $M=90$ methods, described in Section \ref{individual_method_sec} and listed in Table \ref{t:ind_methods_detail} in Appendix \ref{sec:90_methods_appendix}, we considered three smaller pools. In selecting methods for these pools, we sought a diverse set of methods that included methods we felt would be most accurate. We emphasise diversity because this is recommended for combinations of forecasts (\citealp{wang2023forecast}). In Table \ref{t:ind_methods_detail}, asterisks indicate the individual methods included in each of the additional three pools. We now describe the construction of these pools.

(i) We obtained a pool of size $M=27$ methods by selecting, from the pool of 90, methods of different types (simplistic, GARCH, CAViaR and CARE), with and without intraday range and RV. To be more specific, we included in the pool: the more widely-used simplistic benchmarks (HS-100, HS-250 and EWMA); the GARCH models with a Gaussian or skewed $t$ distribution, to reflect one simple and one sophisticated choice of distribution; and the SAV and AS forms of CAViaR and CARE models, with multiplicative ES formulation used in the CAViaR-based models, as we feel these are the most commonly used formulations of the CARE and CAViaR-based approaches.

(ii) We formed a pool of size $M=14$ by removing methods from the pool of 27. We removed the GARCH models with Gaussian distribution, as they are similar to, but less accurate than, the same models with skewed $t$ distribution. We omitted models with intraday range, as we had models in the pool with RV, and the range and RV are both measures of intraday volatility.

(iii) Our smallest pool consisted of a subset of $M=6$ methods from the pool of 14. We included two commonly-used simplistic benchmarks (HS-250 and EWMA). In terms of GARCH models, we included a relatively simple model (GARCH-Return-skewt) and a more sophisticated model (GJR-RV-skewt). We chose a widely-used CAViaR-based model without intraday range or RV (CAViaR-Return-AS-Multplicative) and a CARE model with RV (CARE-RV-AS).

In view of the relatively consistent performance of the combining methods across the five scores in Table \ref{t:Skill_Rank_1800}, in reporting the results for the different pool sizes, in addition to averaging across the six stock indices, we also average over the skill scores for the five different scores. The resulting averaged skill scores are presented in Table \ref{t:PoolSize_Avg_Skill_Rank_1800}, along with ranks also averaged across scores and stock indices. A first comment to make regarding Table \ref{t:PoolSize_Avg_Skill_Rank_1800} is that the relative performances of the combining methods are quite similar for the four different pool sizes, with the best results produced by higher trimmed mean, probability averaging, the performance-based methods and the simple average of all combinations. In terms of how the size of the pool affects the average skill scores for a given method, the table shows that, as the pool size reduces, there is a small drop in performance for the simple average, the median, and almost all the methods based on interval forecast combining. However, for the performance-based weighted combining methods and dynamic selection, the table shows that the best results are achieved with pool sizes of $M=$ 14 and 6. Indeed, we note that the greatest accuracy across all methods and pool sizes corresponds to the minimum score method for $M=$ 14 and 6.

\setlength{\tabcolsep}{5pt}
\begin{table}[ht!]
\captionsetup{font=small}
\centering
\small
\renewcommand{\arraystretch}{1.2}
	\caption{Scoring function results for combining methods applied to the pools of $M=90$, 27, 14 and 6 individual methods. Skill scores and ranks averaged across the six stock indices and five scores. Higher skill scores and lower rank values are better, with grey shading highlighting the three best performing methods in each column. 1800 days used for estimation of individual methods and combining methods.\\}
	\label{t:PoolSize_Avg_Skill_Rank_1800}
	\centering
    \vspace{-10pt}
	\begin{tabular}{lcccc|cccc}\toprule
\multicolumn{1}{c}{} & \multicolumn{4}{c}{Skill score} \vline & \multicolumn{4}{c}{Rank} \\		
Method&$M$=90&$M$=27&$M$=14&$M$=6&$M$=90&$M$=27&$M$=14&$M$=6\\  \midrule
\multicolumn{4}{l}{\textit{Benchmark method}} &\\
\textit{  } Dynamic selection&13.5&13.7&14.3&14.5&   14.9&15.2&11.8&12.1
 \\
\arrayrulecolor{lightgray}\hline
\multicolumn{4}{l}{\textit{Methods based on measures of central tendency}} &\\
\textit{  } Simple average&14.1&13.8&14.0&13.6&9.2&8.6&9.6&11.0
\\
\textit{  } Median&14.2&13.7&14.0&13.9&9.4&11.6&12.2&10.3
\\
\textit{  } Mode&13.9&13.7&13.9&14.0&13.1&12.3&13.4&12.5
\\
\arrayrulecolor{lightgray}\hline
  \multicolumn{4}{l}{\textit{Methods based on interval forecast combining}} &\\
\textit{  } Symmetric trimmed mean&14.0&13.8&14.0&13.8&11.8&10.7&11.3&12.4
\\
\textit{  } Exterior trimmed mean&14.1&13.8&13.9&13.6&10.4&10.8&10.8&12.4
\\
\textit{  } Interior trimmed mean&13.9&13.6&13.7&13.4& 11.5&10.9&12.9&12.1
\\
\textit{  } Lower trimmed mean&13.8&13.7&13.8&13.4&13.0&11.2&12.8&13.7
\\
\textit{  } Higher trimmed mean&14.3 & \cg{14.3} & 14.3 & 14.1 & \cg{6.9} &\cg{5.6}&\cg{6.7}&\cg{7.5}
\\
\textit{  } Flexible trimmed mean&13.8&14.0&13.8&13.9&10.9& \cg{7.5}&10.8&8.6
\\
\textit{  } Probability average& \cg{14.4} & 14.1 &	 14.2 &	13.8 &\cg{6.1}&\cg{6.5}&\cg{6.9}&9.7
\\
\textit{  } Halfspace deepest&14.0&13.6&13.6&13.8& 12.7&14.2&11.8&10.8
\\
\textit{  } Simplicial  deepest&14.0&13.6&13.5&13.9& 12.2&13.9&13.8&10.5
\\
 \arrayrulecolor{lightgray}\hline
  \multicolumn{4}{l}{\textit{Performance-based weighted combining}} &\\
\textit{  } Relative score&\cg{14.4}&\cg{14.3}& \cg{14.6} &\cg{14.7}&10.7&10.5&10.8&10.0
\\
\textit{  } Relative score with weighted median&14.1&\cg{14.3}&14.2&14.2&13.0&11.6&12.9&12.8
\\
\textit{  } Minimum score&\cg{14.5}&\cg{14.6}&\cg{14.9}&\cg{14.9}& 8.9&8.6&\cg{6.4}&\cg{6.9}
\\
\textit{  } Minimum score with ratio&\cg{14.4}&\cg{14.6}&\cg{14.8}& \cg{14.8}&9.7&9.0&8.5&7.8 \\
\textit{  } Minimum score with regularisation&14.3&14.2& 14.4 &14.6& 7.5 & 11.6 & 8.9 &9.7

\\
\arrayrulecolor{lightgray}\hline
  \multicolumn{4}{l}{\textit{Combinations of forecasts produced by combining methods}} &\\
\textit{  } Smooth transition combining&14.3&14.2&14.5&14.5& 10.8&10.8&10.7&11.7
\\
\textit{  } Simple average of all combinations&\cg{14.5}&14.2&14.4&14.4& \cg{7.4}&7.8 & 7.0 & \cg{5.9}
\\
 \arrayrulecolor{black}\bottomrule

\end{tabular}
\end{table}

\subsection{Reduced estimation sample size }\label{reduced_sample_sec}

As we described in Section \ref{Data_sec}, the analysis that we have discussed so far used a rolling window of 1800 days to estimate individual method parameters, and the same number of observations in a rolling window for the estimation of the combining method parameters. We were curious as to whether the relative performance of the combining methods would be affected by radically reducing the estimation window lengths. To investigate this, we repeated the empirical analysis using rolling windows of lengths 900 days in our estimation of both the individual methods and combining methods. This led to the results in Table \ref{t:PoolSize_Avg_Skill_Rank_900}, which is analogous to Table \ref{t:PoolSize_Avg_Skill_Rank_1800} from Section \ref{reduced_pool_sec}. Note that we can compare the skill scores in the two tables because the scores were calculated using the same 1800-day out-of-sample period, and the same benchmark method (HS-250) was used in computing the skill scores. In view of this, it is worth noting that the skill scores are higher for all combining methods and pool sizes in Table \ref{t:PoolSize_Avg_Skill_Rank_900}, indicating that greater accuracy was achieved when using the shorter estimation window lengths of just 900 days.

In terms of the relative performances of the methods for the shorter estimation window length, Table \ref{t:PoolSize_Avg_Skill_Rank_900} shows that, as with the longer estimation window length, performance-based weighted combining produced strong results. Comparing Tables \ref{t:PoolSize_Avg_Skill_Rank_1800} and \ref{t:PoolSize_Avg_Skill_Rank_900}, we see that the superiority of these methods was greater when the smaller estimation window length was used, and that the best results were produced with the smallest of the pool sizes. By contrast, probability averaging performed very well with pool size of $M=90$, but was not competitive for the smaller pool sizes. For the different performance-based weighted combining methods, Tables \ref{t:PoolSize_Avg_Skill_Rank_1800} and \ref{t:PoolSize_Avg_Skill_Rank_900} show that relative score combining becomes more accurate than the more highly-parameterised minimum score methods when the estimation sample size is small.

\setlength{\tabcolsep}{5pt}
\begin{table}[ht!]
\captionsetup{font=small}
\centering
\small
\renewcommand{\arraystretch}{1.2}
	\caption{Scoring function results for combining methods applied to the pools of $M=90$, 27, 14 and 6 individual methods, with just 900 periods used for estimation of individual methods and combining methods. Skill scores and ranks averaged across the six stock indices and five scores. Higher skill scores and lower rank values are better, with grey shading highlighting the three best performing methods in each column.\\}
	\label{t:PoolSize_Avg_Skill_Rank_900}
	\centering
    \vspace{-10pt}
	\begin{tabular}{lcccc|cccc}\toprule
\multicolumn{1}{c}{} & \multicolumn{4}{c}{Skill score} \vline & \multicolumn{4}{c}{Rank} \\		
Method&$M$=90&$M$=27&$M$=14&$M$=6&$M$=90&$M$=27&$M$=14&$M$=6\\  \midrule
\multicolumn{4}{l}{\textit{Benchmark method}} &\\
\textit{  } Dynamic selection&14.9&16.1&\cg{16.8}&17.1&   11.0&9.6&7.4&6.5
 \\
\arrayrulecolor{lightgray}\hline
\multicolumn{4}{l}{\textit{Methods based on measures of central tendency}} &\\
\textit{  } Simple average&15.0&15.5&15.7&15.2&9.9&9.6&10.2&11.3
\\
\textit{  } Median&15.1&15.7&15.7&15.8&8.8&10.1&10.9&11.8
\\
\textit{  } Mode&14.9&15.4&15.5&15.7&12.3&13.4&12.5&13.6
\\
\arrayrulecolor{lightgray}\hline
  \multicolumn{4}{l}{\textit{Methods based on interval forecast combining}} &\\
\textit{  } Symmetric trimmed mean&15.1&15.7&15.8&15.8&9.6&10.3&10.4&11.1
\\
\textit{  } Exterior trimmed mean&15.0&15.6&15.7&15.6&10.8&11.0&11.0&12.2
\\
\textit{  } Interior trimmed mean&14.6&15.2&15.3&14.6& 13.1&12.9&12.7&15.1
\\
\textit{  } Lower trimmed mean&14.7&15.1&15.1&14.9&14.9&15.1&16.0&14.5
\\
\textit{  } Higher trimmed mean&15.1 & 15.6 & 15.4 & 15.3 & 11.3 &10.8&13.8&13.8
\\
\textit{  } Flexible trimmed mean&14.6&15.2&14.8&15.0&15.7& 14.4&17.5&17.1
\\
\textit{  } Probability average& \cg{16.0} & 15.5 & 15.4 & 15.4 &\cg{7.4}&10.9&13.1&12.4
\\
\textit{  } Halfspace deepest&15.1&15.3&15.5&15.2& 11.4&11.1&13.3&13.8
\\
\textit{  } Simplicial  deepest&15.1&15.4&15.7&15.4& 10.7&10.0&12.3&12.2
\\
 \arrayrulecolor{lightgray}\hline
  \multicolumn{4}{l}{\textit{Performance-based weighted combining}} &\\
\textit{  } Relative score&\cg{15.7}&\cg{16.7}& \cg{16.9} &\cg{17.4}&\cg{7.2}&\cg{5.4}&\cg{5.3}&\cg{5.5}
\\
\textit{  } Relative score with weighted median&15.4&16.3&16.5&\cg{17.4}&12.6&10.2&8.5&6.5
\\
\textit{  } Minimum score&\cg{15.7}&16.2&16.4&17.3& 9.7&11.6&8.8&6.5
\\
\textit{  } Minimum score with ratio&15.6&16.2&16.5& 17.2&9.7&11.8&8.7&8.2 \\
\textit{  } Minimum score with regularisation&15.2&\cg{16.5}& 16.6 &17.1& 8.2 & 9.0 & 5.7 &7.0

\\
\arrayrulecolor{lightgray}\hline
  \multicolumn{4}{l}{\textit{Combinations of forecasts produced by combining methods}} &\\
\textit{  } Smooth transition combining&15.4&\cg{16.4}&\cg{16.9}&\cg{17.4}& 9.8&\cg{5.8}&\cg{5.4}&\cg{4.8}
\\
\textit{  } Simple average of all combinations&\cg{15.8}&16.2&16.3&16.7& \cg{5.9}&\cg{6.9} & \cg{6.5} & \cg{6.0}
\\
 \arrayrulecolor{black}\bottomrule

\end{tabular}
\end{table}

\section{Summary and conclusion}\label{sec:summary}

In this study, we explored the effectiveness of combining sizeable pools of VaR and ES forecasts, motivated by the availability of a wide variety of forecasting methods in financial risk management. Using pools of up to 90 individual methods, we evaluated a comprehensive set of combining techniques. In addition to two performance-based weighted methods previously proposed in the literature for a small pool of forecasts, we implemented a set of new combining approaches. These included methods based on interval forecast combining from the decision analysis literature, performance-based weighting with regularisation towards equal weights, and smooth transition combining, which dynamically blends different combining schemes.

In our empirical analysis of six series of stock index returns, we first considered a pool of 90 methods, with estimation based on rolling windows of length 1800 days. Among the most accurate combining methods were a trimmed mean approach and probability averaging. Both of these methods provide VaR and ES forecasts that are more extreme than the simple average, indicating that the large pool of forecasts tended to underestimate risk. Strong results were also produced by performance-based weighted combinations, including the previously proposed relative score and minimum variance methods.

We then considered smaller pools of methods that were subjectively selected, prioritising diversity and accuracy. The most noticeable effect of reducing the pool size was improved accuracy for the performance-based weighted combinations. We also found that shortening the length of the rolling window used for parameter estimation to 900 days delivered improved accuracy for all combining methods, particularly for the performance-based weighted combinations.

The aim of this paper was essentially to challenge the previously proposed relative score and minimum score combining methods by developing alternative combining methods, and investigating the impact on accuracy of applying the methods to a large pool of individual methods. Although our results show potential for several of the new combining methods, our overall finding is that relative score and minimum score combining are hard to beat.

From a managerial and regulatory perspective, our findings support the adoption of forecast combination frameworks as robust tools for improving the reliability of risk measures used in capital allocation, compliance, and strategic financial decision-making.
\newline

\bibliographystyle{chicago}
\bibliography{bibliography}


\newpage
\clearpage
\appendixtitleon
\appendixtitletocon
\begin{appendices}

{\section{Individual forecasting methods}\label{sec:90_methods_appendix}
\par
}

\begin{table}[ht!]
\captionsetup{font=small}
\caption{\label{t:ind_methods_detail} The individual forecasting methods included in the pool of size 90. Methods indicated by * were included in the pools of size 90 and 27. Methods indicated by ** were included in the pools of size 90, 27 and 14. Methods indicated by *** were included in pools of size 90, 27, 14 and 6.}
\vspace{-10pt}
\renewcommand{\arraystretch}{1.2}
\scriptsize
\begin{center}
\begin{tabular}{lll|lll}
\toprule
No.&Method&Type&No.&Method&Type\\ \midrule
1&HS-100\textsuperscript{*}&Nonparametric&46&GJR-GARCH-Range-Gaussian\textsuperscript{*}&Parametric\\
2&HS-250\textsuperscript{***}&Nonparametric&47&GJR-GARCH-Range-t&Parametric\\
3&HS-500&Nonparametric&48&GJR-GARCH-Range-skewt\textsuperscript{*}&Parametric\\
4&HS-1800&Nonparametric&49&GJR-GARCH-Range-Gaussian-EVT&Semi-parametric\\
5&Gaussian-100&Parametric&50&GJR-GARCH-Range-t-EVT&Semi-parametric\\
6&Gaussian-250&Parametric&51&GJR-GARCH-Range-skewt-EVT&Semi-parametric\\
7&Gaussian-500&Parametric&52&GJR-GARCH-Range-Gaussian-FHS&Semi-parametric\\
8&Gaussian-1800&Parametric&53&GJR-GARCH-Range-t-FHS&Semi-parametric\\
9&EWMA\textsuperscript{***}&Parametric&54&GJR-GARCH-Range-skewt-FHS&Semi-parametric\\
10&GARCH-Gaussian\textsuperscript{*}&Parametric&55&GJR-GARCH-RV-Gaussian\textsuperscript{*}&Parametric\\
11&GARCH-t&Parametric&56&GJR-GARCH-RV-t&Parametric\\
12&GARCH-skewt\textsuperscript{***}&Parametric&57&GJR-GARCH-RV-skewt\textsuperscript{***}&Parametric\\
13&GARCH-Gaussian-EVT&Semi-parametric&58&GJR-GARCH-RV-Gaussian-EVT&Semi-parametric\\
14&GARCH-t-EVT&Semi-parametric&59&GJR-GARCH-RV-t-EVT&Semi-parametric\\
15&GARCH-skewt-EVT&Semi-parametric&60&GJR-GARCH-RV-skewt-EVT&Semi-parametric\\
16&GARCH-Gaussian-FHS&Semi-parametric&61&GJR-GARCH-RV-Gaussian-FHS&Semi-parametric\\
17&GARCH-t-FHS&Semi-parametric&62&GJR-GARCH-RV-t-FHS&Semi-parametric\\
18&GARCH-skewt-FHS&Semi-parametric&63&GJR-GARCH-RV-skewt-FHS&Semi-parametric\\
19&GARCH-Range-Gaussian\textsuperscript{*}&Parametric&64&CAViaR-SAV-Multplicative\textsuperscript{**}&Semi-parametric\\
20&GARCH-Range-t&Parametric&65&CAViaR-SAV-Additive&Semi-parametric\\
21&GARCH-Range-skewt\textsuperscript{*}&Parametric&66&CAViaR-AS-Multplicative\textsuperscript{***}&Semi-parametric\\
22&GARCH-Range-Gaussian-EVT&Semi-parametric&67&CAViaR-AS-Additive&Semi-parametric\\
23&GARCH-Range-t-EVT&Semi-parametric&68&CAViaR-IG-Multplicative&Semi-parametric\\
24&GARCH-Range-skewt-EVT&Semi-parametric&69&CAViaR-IG-Additive&Semi-parametric\\
25&GARCH-Range-Gaussian-FHS&Semi-parametric&70&CAViaR-Range-SAV-Multplicative\textsuperscript{*}&Semi-parametric\\
26&GARCH-Range-t-FHS&Semi-parametric&71&CAViaR-Range-SAV-Additive&Semi-parametric\\
27&GARCH-Range-skewt-FHS&Semi-parametric&72&CAViaR-Range-AS-Multplicative\textsuperscript{*}&Semi-parametric\\
28&GARCH-RV-Gaussian\textsuperscript{*}&Parametric&73&CAViaR-Range-AS-Additive&Semi-parametric\\
29&GARCH-RV-t&Parametric&74&CAViaR-Range-IG-Multplicative&Semi-parametric\\
30&GARCH-RV-skewt\textsuperscript{**}&Parametric&75&CAViaR-Range-IG-Additive&Semi-parametric\\
31&GARCH-RV-Gaussian-EVT&Semi-parametric&76&CAViaR-RV-SAV-Multplicative\textsuperscript{**}&Semi-parametric\\
32&GARCH-RV-t-EVT&Semi-parametric&77&CAViaR-RV-SAV-Additive&Semi-parametric\\
33&GARCH-RV-skewt-EVT&Semi-parametric&78&CAViaR-RV-AS-Multplicative\textsuperscript{**}&Semi-parametric\\
34&GARCH-RV-Gaussian-FHS&Semi-parametric&79&CAViaR-RV-AS-Additive&Semi-parametric\\
35&GARCH-RV-t-FHS&Semi-parametric&80&CAViaR-RV-IG-Multplicative&Semi-parametric\\
36&GARCH-RV-skewt-FHS&Semi-parametric&81&CAViaR-RV-IG-Additive&Semi-parametric\\
37&GJR-GARCH-Gaussian\textsuperscript{*}&Parametric&82&CARE-SAV\textsuperscript{**}&Semi-parametric\\
38&GJR-GARCH-t&Parametric&83&CARE-AS\textsuperscript{**}&Semi-parametric\\
39&GJR-GARCH-skewt\textsuperscript{**}&Parametric&84&CARE-IG&Semi-parametric\\
40&GJR-GARCH-Gaussian-EVT&Semi-parametric&85&CARE-Range-SAV\textsuperscript{*}&Semi-parametric\\
41&GJR-GARCH-t-EVT&Semi-parametric&86&CARE-Range-AS\textsuperscript{*}&Semi-parametric\\
42&GJR-GARCH-skewt-EVT&Semi-parametric&87&CARE-Range-IG&Semi-parametric\\
43&GJR-GARCH-Gaussian-FHS&Semi-parametric&88&CARE-RV-SAV\textsuperscript{**}&Semi-parametric\\
44&GJR-GARCH-t-FHS&Semi-parametric&89&CARE-RV-AS\textsuperscript{***}&Semi-parametric\\
45&GJR-GARCH-skewt-FHS&Semi-parametric&90&CARE-RV-IG&Semi-parametric\\
\bottomrule

\end{tabular}

\end{center}

\end{table}

\section{Probability averaging implementation}\label{sec:prob_averaging_implementation}

For the probability averaging approach, the detailed implementation procedure to infer a CDF from the VaR and ES produced by the nonparametric and semi-parametric individual methods is shown below:
\begin{itemize}
\item We generate three parameter grids: $300$ equally spaced trial standard deviation values $\sigma \in (0, 10]$, $100$ equally spaced trial degrees of freedom values $\nu \in (2, 30]$, and $200$ equally spaced trial skewness values $\lambda \in (-1, 1)$. The ranges for each parameter are chosen based on the empirical estimation results from the parametric GARCH-skew-$t$ model.

\item Following the calculation of VaR and ES for the skew-$t$ distribution, we generate candidate VaR and ES values over a parameter grid of size $300 \times 100 \times 200$. Each pair of VaR and ES values corresponds to a unique combination of skew-$t$ distribution parameters.

\item For each model and forecast origin, we search over trial values of $\sigma$, $\nu$, and $\lambda$ for each pair of VaR and ES forecasts, and retain the parameter sets for which the average squared distance between the matched and actual VaR and ES forecasts is below 0.001. Among the retained sets, we select the combination of $\sigma$, $\nu$, and $\lambda$ with the lowest skewness, i.e., the $\lambda$ value closest to zero. Figure 5 illustrates the 90 CDFs corresponding to the 90 VaR and ES forecast pairs at the final forecast origin in the FTSE 100 analysis.

\item For each of the 1000 equally spaced values on the $x$-axis generated in range $[-20,20]$, the corresponding 90 CDF values are averaged to produce the single combined CDF, as demonstrated by the black line in Figure 5 in the paper. The range of the $x$-axis only displays $[-5,5]$ for clarity of the presentation.

\item The combined VaR is computed from this CDF via interpolation, identifying the smallest $x$-axis value where its corresponding CDF reaches or exceeds the probability level $\alpha=2.5\%$, as demonstrated in Figure 5 in the paper.

\item For ES, defined as $ES = \frac{1}{\alpha} \int_{-\infty}^{\mathrm{VaR}} x \, dF(x)$, we approximate it using a discrete representation of the probability density function (PDF) and numerical integration. First, we identify the $x$-values that are less than or equal to the combined VaR and record their corresponding CDF values. The PDF is then approximated by taking the differences between consecutive CDF values and multiplying them by the midpoints of the corresponding consecutive $x$-value pairs. These products are summed and divided by the probability level \((\alpha = 2.5\%)\) to obtain the combined ES forecast.

\end{itemize}

\end{appendices}
\end{document}